# Effect of Stress and Surface Roughness on Electrodeposition in All-Solid-State Batteries: A Computational Investigation

**Kaniza Islam**[1], **Ayush Morchhale**[1], **Jung-Hyun Kim**[1], **Yanzhou Ji**[2,*], **Noriko Katsube**[1]

[1] *Department of Mechanical and Aerospace Engineering, The Ohio State University, Columbus, OH, 43210, United States*

[2] *Department of Materials Science and Engineering, The Ohio State University, Columbus, OH, 43210, United States*

## Abstract

All-solid-state batteries (ASSBs) promise high energy density and enhanced safety, but their development is hindered by instability and incompatibility at solid-solid interfaces. In Li-metal ASSBs, lithium penetration occurs despite stiff ceramic electrolytes via grain boundaries, often initiated by minor Li/SE interfacial irregularities. Here we introduce a two-dimensional continuum model with electro-chemo-mechanical coupling to investigate interfacial current distribution in Li ASSBs with surface-roughened argyrodite electrolyte under stack pressures and applied current density. Our theoretical analysis and simulation studies highlight the critical role of mechanical stress in interfacial current distribution. We find that prominent stress variations around elongated surface protrusions are the key to nonuniform Li deposition, without which Li deposition becomes uniform even on a rough surface. Moreover, our parametric study elucidates that stress effects dominate the overpotential and current distribution under low interfacial current density to exchange current density ratios, otherwise the high interfacial resistance due to surface-roughness-induced interfacial area becomes dominant. With these insights, we also discuss the potential of engineering artificial interlayers to modulate interfacial current distributions, offering guidance for improving the long-term performance and reliability of ASSBs.

## 1. Introduction

All-solid-state batteries (ASSBs) have emerged as a breakthrough energy storage technology, offering significant advantages over conventional lithium-ion batteries (LIBs). The transition to solid-state systems replaces flammable liquid electrolytes with nonflammable inorganic solid electrolytes (SEs), thereby eliminating leakage and combustion risks while mitigating short-circuiting caused by dendrite penetration. Moreover, ASSBs offer faster charging capability. In particular, the use of Li metal anodes in ASSBs enables a significant increase in volumetric capacity, leading to a substantial improvement in energy density. Owing to these advantages, ASSBs have strong potential in next-generation, high-performance, and safer energy storage systems [1-3].

Despite these advantages, ASSBs face critical challenges related to the electrochemical and mechanical stability of the Li/SE interfaces [4]. Notably, although the free growth of mossy Li dendrites at the anode/electrolyte interfaces can be largely suppressed in ASSBs due to mechanical constraints imposed by external pressure, the Li penetration cannot be fully eliminated, which is often triggered by minor surface irregularities at the Li/SE interface and facilitated by pre-existing microstructure defects in SEs such as grain boundaries (GBs), voids, or microcracks as illustrated in Figure 1. Li penetration into SEs can induce crack initiation and propagation, thereby facilitating further Li penetration and adversely affecting long-term battery safety. The resulting localized current concentration promotes nonuniform Li deposition. This feedback accelerates dendrite growth, leading to mechanical degradation, internal short circuits, and reduced cycle life and safety [5-8]. Meanwhile, during discharging of ASSBs, the nonuniform stripping of Li can lead to the formation of micropores at Li/SE interface, which increases interfacial impedance and reduces ionic conductivity [9-11]. ASSBs remain vulnerable to dendrite and void formation during high-rate Li plating and stripping [9-12], regardless of SE composition. These interfacial instabilities arise from the coupled effects of ion transport, interfacial electrochemistry, and mechanical stress. Experimental studies have revealed dendrite-induced short circuits in various SE materials [13-17]; however, they remain limited in capturing dynamic dendrite evolution or quantifying local stress, current density, and potential distributions. Consequently, the underlying mechanisms governing Li penetration and interface degradation are not yet fully understood.

* Email: ji.730@osu.edu

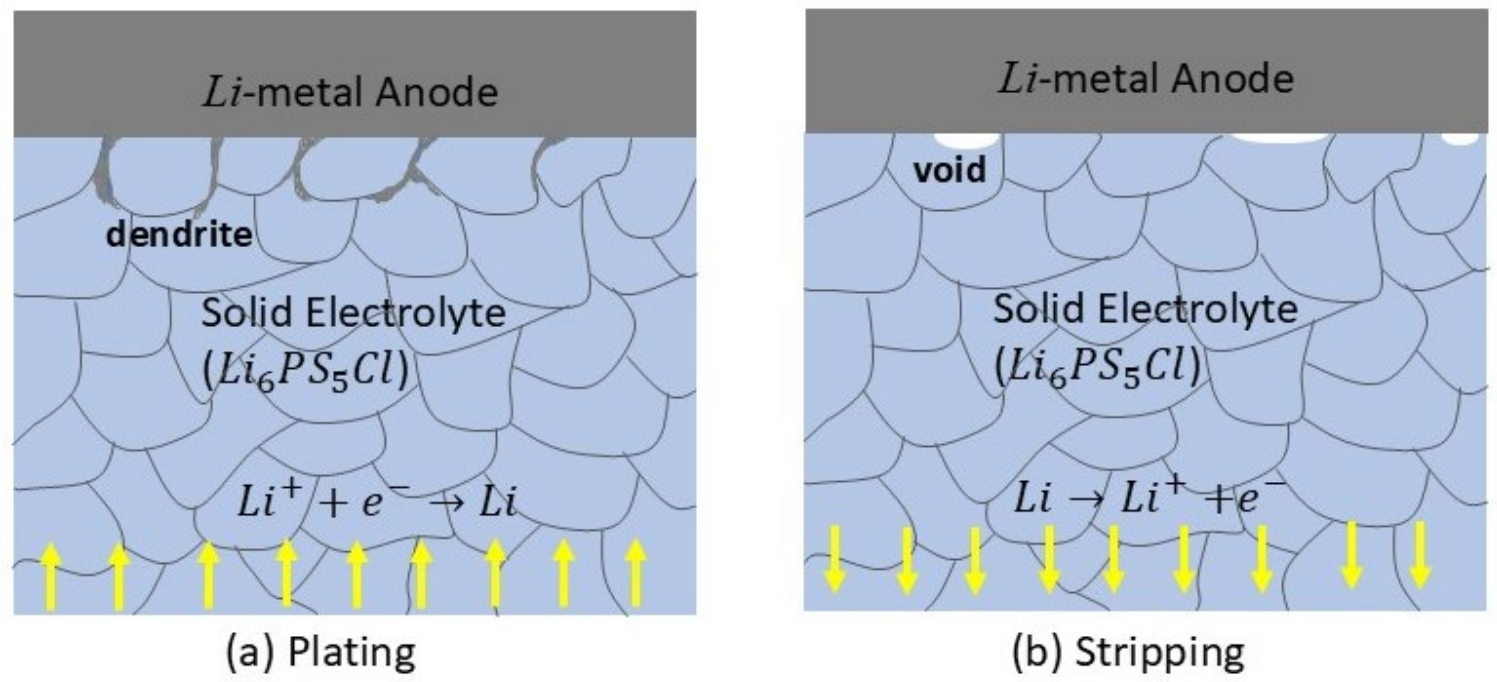

**Fig. 1** Schematic illustrations of dendrite and void formation during (a) plating and (b) stripping in ASSBs, where the yellow arrows represent the direction of the applied current density and the thin gray curves represent grain boundaries (GBs) in the electrolyte.

Modeling approaches have become essential for probing the complex Li/SE interfacial phenomena and the related physical/chemical fields that are difficult to measure experimentally. Since these interfacial phenomena involves multiple coupled physical/chemical processes such as electrochemical reactions for Li deposition (during plating) and dissolution (during stripping), transport of $Li^+$ and electrons, and mechanical interactions between Li and SE, a predictive modeling framework must therefore be developed with advanced electro-chemo-mechanical coupling capable of resolving local stress concentrations, defect-driven deposition, and geometry-sensitive current distributions critical to dendrite suppression. Early modeling studies by Monroe and Newman [18,19] showed that dendrite suppression requires SE shear modulus at least twice that of Li metal, with interfacial stress and surface overpotential governing current distribution. Extended models incorporating interfacial geometry, stress, material nonlinearity, and microstructural effects revealed strong coupling in current distribution [20-24], yet experiments show that Li penetration can still occur despite meeting the modulus criterion, primarily due to surface defect morphology [14-17,25]. To capture the subsequent morphological evolution of Li dendrites, phase-field modeling (PFM) [26-29] has attracted increasing interests. Although PFM of Li/SE interface evolution is our long-term goal, this study will primarily focus on static stress and current density distributions of given Li/SE geometry. Such static investigations can already provide useful insights into Li electrodeposition behaviors. Therefore, we explicitly model the interfacial kinetics through a modified Butler-Volmer equation rather than a phase-field formulation. Fracture-mechanics and continuum models incorporating stress-dependent Butler-Volmer kinetics show that stress-current coupling governs Li plating and crack evolution, with surface or material inhomogeneities driving nonuniform deposition and lower exchange current density promoting more uniform Li growth [17, 30-37].

Stable Li deposition is governed by the interplay among mechanical stress, interfacial geometry, and electrochemical kinetics, and these factors have been investigated through a range of experimental and theoretical approaches. Tu et al. [35] used a continuum model incorporating conformal interfacial surface defects to investigate the evolution of interfacial morphology during cycling of Li deposition. Their results showed that current preferentially concentrates near defects, producing a nonuniform normal current density distribution. They concluded that stable Li deposition requires a lower exchange current density and a wide, shallow interfacial surface. Similarly, Verma et al. [36] developed a model assuming perfect conformal contact at the Li/SE interface, where interfacial surface roughness was represented by a sinusoidal perturbation. Their results showed that increasing external pressure enhances deposition stability when the shear modulus ratio is low. In contrast, deposition instabilities become more pronounced with increasing applied current density, as the resulting stress is insufficient to redirect Li flux away from growing Li protrusions. Furthermore, greater surface roughness at the metal/electrolyte interface promotes transport-driven instabilities. The study emphasized that maintaining a smooth interface and high ionic conductivity is critical for achieving stable deposition and good performance in ASSBs. Mechanical stress has also been shown to directly alter the thermodynamics and kinetics of Li deposition and dissolution at the Li/SE interface. Key strategies explored to improve deposition stability include the use of artificial interlayers [38-40], application of external pressure to reduce interfacial resistance [38,40-43], increasing SE density (>95%) [44], high-shear-modulus alloy (HSEA) pairing of alloy anodes and electrolytes [45], optimization of SE composition,[44] adoption of anode-free architectures with buffer layers between the current collector and SE [47,48], and interface engineering approaches [49-53].

However, the combined effects of stack pressure, protruded interfacial geometry, and exchange current density on deposition stability remain insufficiently explored even in static analyses without interface evolution. Previous studies did not fully explore how variations in protruded interfacial morphology, compared with a smooth interface, influence overpotential and Li deposition behaviors under mechanical loading, applied current density, and variations of exchange current density. As a result, it is often concluded that the stress effect is negligible. A key distinction of the present work from previous studies is that the mechanical stress contribution can be selectively included and excluded in the simulations, allowing its influence on interfacial kinetics to be clarified directly. Addressing these gaps would enable more quantitative and predictive guidance for tuning stack pressure, controlling interfacial roughness, and selecting SE or interlayer materials to mitigate unstable Li deposition and dendrite formation.

This study aims to computationally investigate the effects of mechanical stress and interfacial surface roughness geometry on metal electrodeposition in metal/inorganic-SE systems employed in ASSBs. A coupled electro-chemo-mechanical modeling framework is used to simulate static stress and current density distribution in Li-metal ASSB systems under realistic operating conditions, incorporating boundary constraints that reflect practical stack pressures. This work adopts a simplified Li/SE interfacial model with several deliberate assumptions. Full electrochemical cycling, time-dependent interface evolution, and continuously distributed surface defects are not considered. In addition, the study does not include dendrite initiation, crack propagation, void growth, contact loss, or morphological evolution during cycling. Our results reveal that mechanical stress is the primary driver of non-uniform Li deposition, even in the presence of surface roughness. The geometry of surface roughness governs local stress variations and promotes preferential Li growth at extruded features. Furthermore, the influence of mechanical stress becomes increasingly dominant at high exchange current densities and low interfacial resistance. These findings indicate that achieving stable Li deposition requires the use of low-resistance interlayers with high ionic conductivity and low elastic modulus. The following sections present our modeling framework and key findings: Section 2 outlines the electro-chemo-mechanical coupling approach, while Section 3 analyzes the impact of interfacial geometry and stack pressure and variation of exchange current and applied current densities along the interface of the surface roughness. Although achieving uniformly stable Li deposition remains challenging, our study demonstrates that careful tuning of electrochemical and mechanical parameters can enhance interface stability and suppress dendritic growth.

## 2. Methods

### *2.1 Model Geometry and Description*

We adopt a simplified Li/SE interfacial model with several deliberate assumptions. Full electrochemical cycling, time-dependent interface evolution, and continuously distributed surface defects are not considered. Although these features better represent real systems, they introduce complex interactions that are beyond the scope of this study. Instead, isolated surface roughness features are assumed, enabling a controlled investigation of how roughness geometry influences interfacial stress, current density distribution, and overpotential.

We consider a half-cell in our simulations. The geometry used is schematically shown in Figure 2. For computation efficiency considerations, the system is reduced to 2D geometry with isotropic properties. Both the Li-metal anode and the argyrodite $Li_6PS_5Cl$ SE (with thickness “L”) are assumed to be single crystals without any defects, and they are perfectly bonded and therefore do not account for interfacial gaps, void formation, microcracks, or grain-boundary transport. In practical systems, such features can introduce additional current localization mechanisms (e.g., contact loss) and may provide preferred pathways for Li penetration. As a result, the stress-induced effects identified here should be interpreted as representative of an idealized interface, and their relative importance may differ in systems with interfacial defects or microstructural heterogeneity.

In reality, minute rough surfaces are formed during the smoothing process or after fine surface polishing of the SE in cell manufacturing, which can gradually contribute to potential dendrite growth. We are guided by experimental SE surface roughness data which is about 100 nm. In our model, the metal-electrolyte surface roughness is represented by various dimensions of a half ellipse as shown in Figure 2(c). The external stack pressure "P" is applied on top of the Li anode. The interfacial distance "s" is measured from the tip of an ellipse as shown in Figure 2(b). The current density denoted by "$i_{app}$" is applied at $y=0$, and the interfacial current density at $y=L$, denoted by "$i_{BV}$" is governed by the interfacial kinetic reaction of Li plating/stripping. The case with a protective $Li_3N$ layer with thickness "t" as shown in Figure 2(d) and Figure 2(e) will also be investigated.

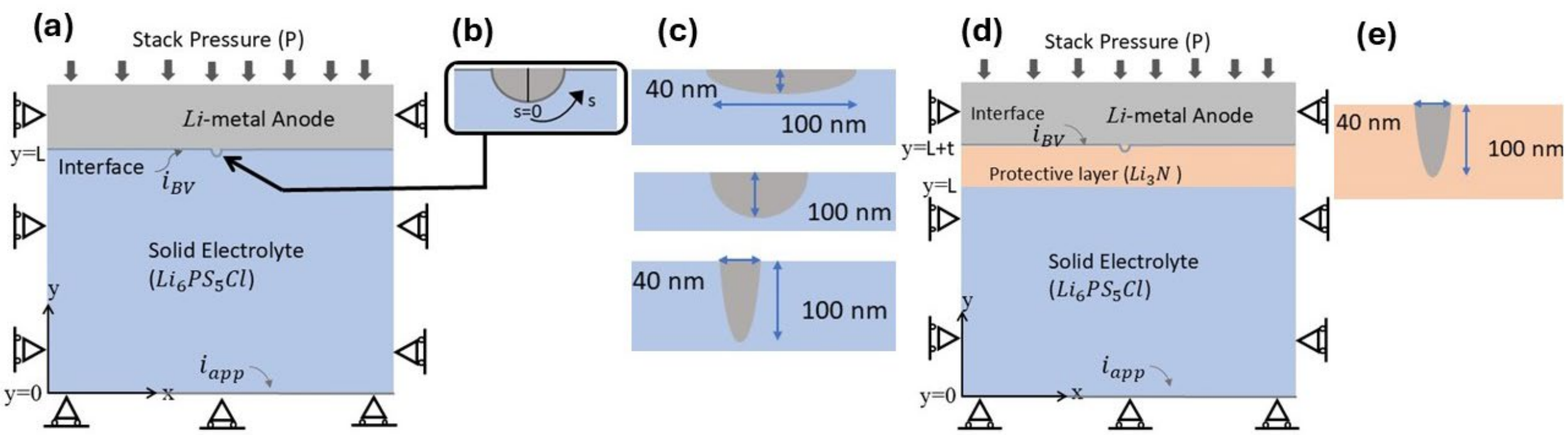

**Fig. 2** Schematic illustrations of **(a)** a 2D half-cell model consisting of a Li metal anode and SE showing interfacial roughness at the anode/SE interface; **(b)** Interface distance measured along the protrusion; **(c)** Various geometries of interfacial roughness; **(d)** 2D half-cell model including the protective layer; **(e)** Elongated ellipse roughness with a protective layer.

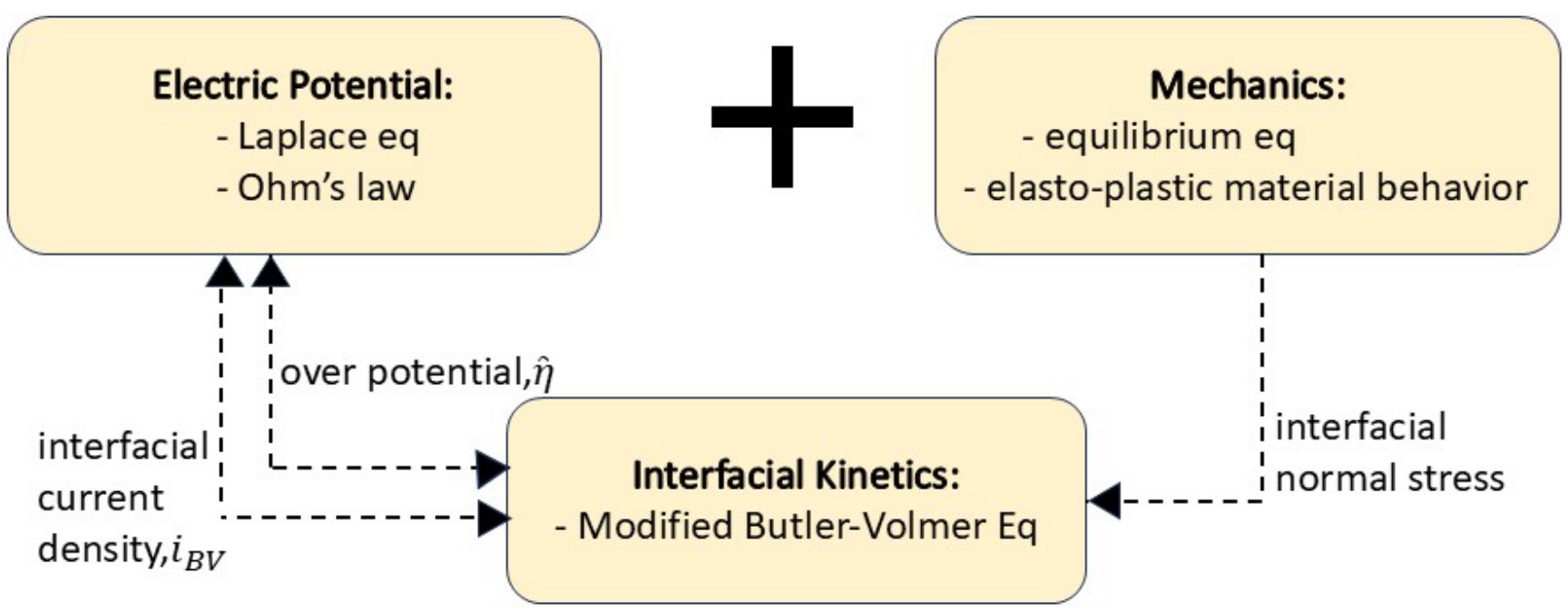

**Fig. 3** Schematic illustration of the coupling strategy of electric effect, mechanical effect and interfacial kinetics in our current model.

The model (Fig. 3) establishes a coupled electro-chemo-mechanical framework, where the electric potential field in SE is governed by the Laplace equation, and the mechanical fields are governed by equilibrium equations and elasto-plastic material behavior in the whole system. These physical fields are interconnected through interfacial kinetics described by a modified Butler-Volmer equation. The coupling between the electric potential field and interfacial kinetics governs the overpotential and the interfacial current density, whereas the coupling with the mechanical field provides the interfacial normal stress influencing the reaction kinetics. This bidirectional coupling captures the mutual influence of electrochemical and mechanical effects at the interface.

### *2.1.1 Ion Conduction (electric effect)*

The bulk SE is assumed to be single crystal without defects. $Li^+$ ions are homogeneously distributed throughout the bulk, resulting in purely ohmic conduction. Assuming electroneutrality, the electric potential distribution within the system is governed by the Laplace equation,

$$\nabla^2\phi_{SE} = 0 \ , i_{SE} = -\sigma_{SE}^{+}\nabla\phi_{SE} \tag{1}$$

Here, ionic conductivity of $Li^+$ is denoted by $\sigma_{SE}^{+}$ and electric potential within SE by $\phi_{SE}$. The boundary conditions are,

$$\text{at } y=0,\ i=i_{app} \text{ and at } y=L,\ i=i_{BV} \text{ for Fig 2(a)} \tag{2}$$

For the case with the protective layer, we adopt similar assumptions. The governing equation in the layer is

$$\nabla^2\phi_{layer} = 0, i_{layer} = -\sigma_{layer}^{+}\nabla\phi_{layer} \tag{3}$$

Where $\sigma_{layer}^{+}$ and $\phi_{layer}$ are the $Li^+$ conductivity and electric potential in the layer respectively. The boundary conditions for Figure 2(d) are given by

$$\text{at } y=0,\ i=i_{app}, \text{ at } y=L+t,\ i=i_{BV} \text{ and at y}=L,\ i_{SE}=i_{layer}, \phi_{SE}=\phi_{layer} \tag{4}$$

### *2.1.2 Elasto-Plastic behavior (mechanical effect)*

Mechanical equilibrium is assumed for both the Li metal and the SE,

$$\nabla\cdot\boldsymbol{\sigma} = 0. \tag{5}$$

During Li deposition, when the effective stress remains below their yield strengths, the materials undergo elastic deformation. In this elastic regime, the stress-strain behavior of Li metal, the SE and the protective layer are described by linear elastic constitutive relations, expressed as follows:

$$\boldsymbol{\sigma} = \frac{E_i}{1+\nu_i}\boldsymbol{\varepsilon} + \frac{\nu_i E_i}{(1+\nu_i)(1-2\nu_i)} trace(\boldsymbol{\varepsilon}) I \tag{6}$$

Here, Young's modulus ($E_i$) and Poisson's ratio ($\nu_i$) for Li metal ($i=Li$), SE ($i=SE$) and protective layer ( $i=layer$). Notably, SE materials commonly used in solid-state batteries tend to have high elastic moduli and low fracture toughness [35] making the assumption of linear elasticity reasonable.

Once the effective stress in any region of the Li metal exceeds its yield strength, the material undergoes plastic deformation, requiring a different stress-strain relationship. For Li metal, an elastic-perfectly plastic model is adopted. The von Mises yield criterion is applied to determine whether the Li metal has entered the plastic deformation regime:

$$\phi(\sigma) = \sqrt{\frac{3}{2}}\left|dev(\boldsymbol{\sigma})\right| - Y_{Li} = 0 \tag{7}$$

The Levy-Mises associated flow rule is employed to characterize the plastic behavior of Li metal during deformation:

$$d\boldsymbol{\varepsilon}^{p} = d\lambda\frac{d\phi}{d\sigma} = \frac{3}{2}J_{2}^{-0.5}d\lambda \cdot \boldsymbol{s} \tag{8}$$

$d\boldsymbol{\varepsilon}^{p}$ is the incremental of plastic strain tensor, $d\lambda$ is the plastic multiplier that can be determined from the constraint of perfect plasticity, $J_2$ is the second invariant of the deviatoric stress "$\boldsymbol{s}$" and $J_2 = \frac{3}{2}\boldsymbol{s}:\boldsymbol{s}$ .

***2.1.3 Interfacial Reaction (electrochemical and mechanical coupling)***

At the interface, the current density during electrodeposition is influenced by surface roughness and can be described by a stress-modified Butler-Volmer equation, which accounts not only for the electrochemical overpotential but also for mechanical stress, reflecting how the mechanical state of the electrode alters the energy landscape of Li plating/stripping [18,19]. The Butler-Volmer equation from McMeeking et al. [62] can be written as,

$$i_{BV} = i_o\left\{\exp\left[\frac{(1-\beta)F\tilde{\eta} - \Omega_{etrode}^{Li}\sigma_n}{RT}\right] - \exp\left[\frac{-\beta F\tilde{\eta}}{RT}\right]\right\} \tag{9}$$

which can also be rearranged as

$$i_{BV} = i_o\exp\left(-\frac{\beta\Omega_{etrode}^{Li}\sigma_n}{RT}\right)\left[\exp\left(\frac{(1-\beta)F\tilde{\eta}'}{RT}\right) - \exp\left(\frac{-\beta F\tilde{\eta}'}{RT}\right)\right] \tag{10}$$

where $i_0$ is the exchange current density, $\tilde{\eta}' = \tilde{\eta} - \frac{\Omega_{etrode}^{Li}\sigma_n}{F}$ is the mechanics modified overpotential, and $\tilde{\eta}$ is the mechanics-free overpotential (which will later be referred as "overpotential") given by

$$\tilde{\eta} = \phi_{Li} - \phi_{SE} - \phi_{eq,SE} \text{ for Figure 2(a)}$$
$$\tilde{\eta} = \phi_{Li} - \phi_{layer} - \phi_{eq,layer} \text{ for Figure 2(d)} \tag{11}$$

Here, $\phi_{Li}$=0V for this study and $\phi_{eq,SE}$ and $\phi_{eq,layer}$ are the equilibrium potential. $\Omega_{etrode}^{Li}$ is the molar volume of Li and $\sigma_n$ is the interfacial normal stress evaluated by

$$\sigma_n = \boldsymbol{n}\cdot(\boldsymbol{\sigma}\cdot\boldsymbol{n}) \tag{12}$$

where $\boldsymbol{n}$ is the unit interfacial vector directing outward from the electrode to the SE/layer.

In general, the mechanical contribution to the chemical potential in solids is governed by hydrostatic stress. However, for interfacial reactions, lithium insertion occurs normal to the interface, and the relevant quantity is the normal traction. Accordingly, following Deshpande and McMeeking (2023) [62], we express the mechanical contribution to the overpotential in terms of $\sigma_n$. In the present 2D setting, stress variations are dominated by the normal component, making this a reasonable approximation. All the physical processes are coupled together and numerically solved using the finite element method using COMSOL (Supplementary Information, S1).

Here we would like to further analyze the governing factors for the mechanics modified overpotential. Using symmetry factor $\beta = 0.5$, Eq. (10) becomes

$$\tilde{\eta}' = \frac{2RT}{F} \ln \left[ \frac{i_{BV}}{2i_o \exp\left( \frac{-\Omega^{Li}_{etrode} \sigma_n}{2RT} \right)} + \sqrt{\left( \frac{i_{BV}}{2i_o \exp\left( \frac{-\Omega^{Li}_{etrode} \sigma_n}{2RT} \right)} \right)^2 + 1} \right] \tag{13}$$

indicating that the normal stress, exchange current density, and applied current density $(\sim i_{BV})$ all affect the overpotential. If the expressions inside the exponential on the right-hand side of Eq. (9) are significantly less than 1, then Eq. (9) can be linearized by a first-order Taylor expansion as:

$$i_{BV} = \frac{\tilde{\eta}'}{R_{\text{int}}} = \frac{1}{R_{\text{int}}} \left( \tilde{\eta} - \frac{\Omega^{Li}_{etrode} \sigma_n}{F} \right), \; R_{\text{int}} = \frac{RT}{i_o F} \tag{14}$$

which leads to

$$\tilde{\eta} = R_{\text{int}} i_{BV} + \frac{\Omega^{Li}_{etrode} \sigma_n}{F} \tag{15}$$

where $R_{\text{int}}$ is the interfacial resistance. Eq. (15) indicates that the electrodeposition behavior could be dominated by mechanical effects under high stack pressures, low interfacial current densities, or high exchange current densities, where the term $\frac{\Omega^{Li}_{etrode} \sigma_n}{F}$ dominates the term $R_{\text{int}} i_{BV}$.

### *2.2 Experimental Procedures for Determining Exchange Current Density*

The exchange current density was experimentally determined by fitting the Tafel plot using Gamry Echem Analyst software (see Supplementary Information S2). Tafel plots (Gamry 600+) were obtained at a scan rate of 1 mV/s within the range of -0.2 to 0.2 V. Symmetric cells were fabricated by pressing 100 mg of $Li_6PS_5Cl$ SE into 10 mm diameter pellets at 700 MPa. Li or $Li_3N$–Li discs (9 mm diameter) were then pressed onto both sides of SE pellet at 150 MPa. A constant stack pressure of 20 MPa was maintained throughout cell operation. The $Li_3N$-Li discs were obtained by single-step facile gas phase reaction between Li-metal and $N_2$ gas using the procedure mentioned by Tang et al. [54,55].

## .Results and Discussion

In this study, we examine the impact of interfacial morphology on ongoing Li deposition and stripping at the interface. Initially, we explore the variations in current density and the stress effects at the Li metal-SE interface, which stems from the initial geometric roughness introduced during the manufacturing process. The Li deposition/stripping is governed by charge transfer reactions at the interface, as described by the stress-modified nonlinear Butler-Volmer relation. The parameters used in this investigation are tabulated in Table I.

**Table I.** Material properties for modeling coupled electro-chemo-mechanical behavior.

| Parameters | Name | Symbol | Value | Unit | Ref |
|---|---|---|---|---|---|
| Mechanical | Young's modulus of **Li** | $E_{Li}$ | 7.8 | GPa | [56] |
| | Poisson's ratio of **Li** | $\nu_{Li}$ | 0.38 | 1 | [56] |
| | Yield strength of **Li** | $Y_{Li}$ | 0.8 | MPa | [56] |
| | Poisson's ratio of $\mathbf{Li_6PS_5Cl}$ | $\nu_{SE}$ | 0.37 | 1 | [56] |
| | Poisson's ratio of $\mathbf{Li_3N}$ | $\nu_{layer}$ | 0.22 | 1 | [61] |
| | Young's modulus of $\mathbf{Li_6PS_5Cl}$ | $E_{SE}$ | 22.1 | GPa | [57] |
| | Young's modulus of $\mathbf{Li_3N}$ | $E_{layer}$ | 48 | GPa | [61] |
| | Molar volume of **Li** | $\Omega_{etrode}^{Li}$ | 1.3 x$10^{-5}$ | $m^3$/mol | [62] |
| | Stack Compressive Stress | P | 3 | MPa | |
| Electrochemical | $Li^+$ conductivity in $\mathbf{Li_6PS_5Cl}$ | $\sigma_{SE}^{+}$ | 0.319 | S/m | [58] |
| | $Li^+$ conductivity in $\mathbf{Li_3N}$ | $\sigma_{layer}^{+}$ | 0.02 | S/m | [59,60] |
| | Applied current density | $i_{app}$ | 100 | $\mu A/cm^2$ | |
| | Exchange current density between $\mathbf{Li/Li_6PS_5Cl}$ | $i_{o,SE}$ | 95 | $\mu A/cm^2$ | Exp |
| | Exchange current density between $\mathbf{Li/Li_3N}$ | $i_{o,layer}$ | 133 | $\mu A/cm^2$ | Exp |
| | Temperature | T | 298 | K | Exp |
| Geometry | $\mathbf{Li_6PS_5Cl}$ thickness | $L$ | 20 | μm | |
| | $\mathbf{Li_3N}$ layer thickness | $t$ | 3 | μm | Exp |

While this simulation employed a yield strength of 0.8 MPa with perfect plasticity for lithium, consistent with previous studies [35], it is important to note that the yield strength of lithium is size-dependent, as demonstrated by Stallard et al. [63]. For small protrusions, the local yield strength of lithium may be higher than the value used in this study. Furthermore, rate-dependent behavior and creep, as documented in the literature [64, 65], were not considered.

### *3.1 Normal Stress Distribution at the Li/SE Interface*

Applying compressive stack pressure is essential during cell operation to maintain intimate contact at the interface. Ideally, this pressure should be sufficient to induce plastic deformation in the Li metal, helping to smooth out inhomogeneities caused by uneven deposition.

Given the boundary conditions shown in Figure 2(a), the equilibrium equation (5) is solved by using mechanical constitutive equations (6), (7) and (8) with the material parameters given in Table I. In Figure 4, the interfacial normal stress distributions for different surface geometries at the Li-metal/SE interface are shown. The normal stress values are evaluated along the interfacial curve of the protrusion with the origin located at the peak of the bump (protrusion) shown in Figure 2(b). For the smooth surface away from the roughness, the interfacial normal stress maintains a constant pressure of 3 MPa regardless of the shape of the roughness. Large jump of stress values indicates the sharp corners, located at the edges of the roughness. The rough surfaces exhibited distinct normal stress distributions depending on the geometry. For both the circular and flat elliptical shapes, changes in normal compressive stress were observed mainly at the edges, decreasing from 3 MPa to approximately 2.6 MPa and 2.9 MPa respectively at the center. Meanwhile, the elongated ellipse has shown the greatest variation, with the compressive stress dropping from 3 MPa to approximately 1.6 MPa.

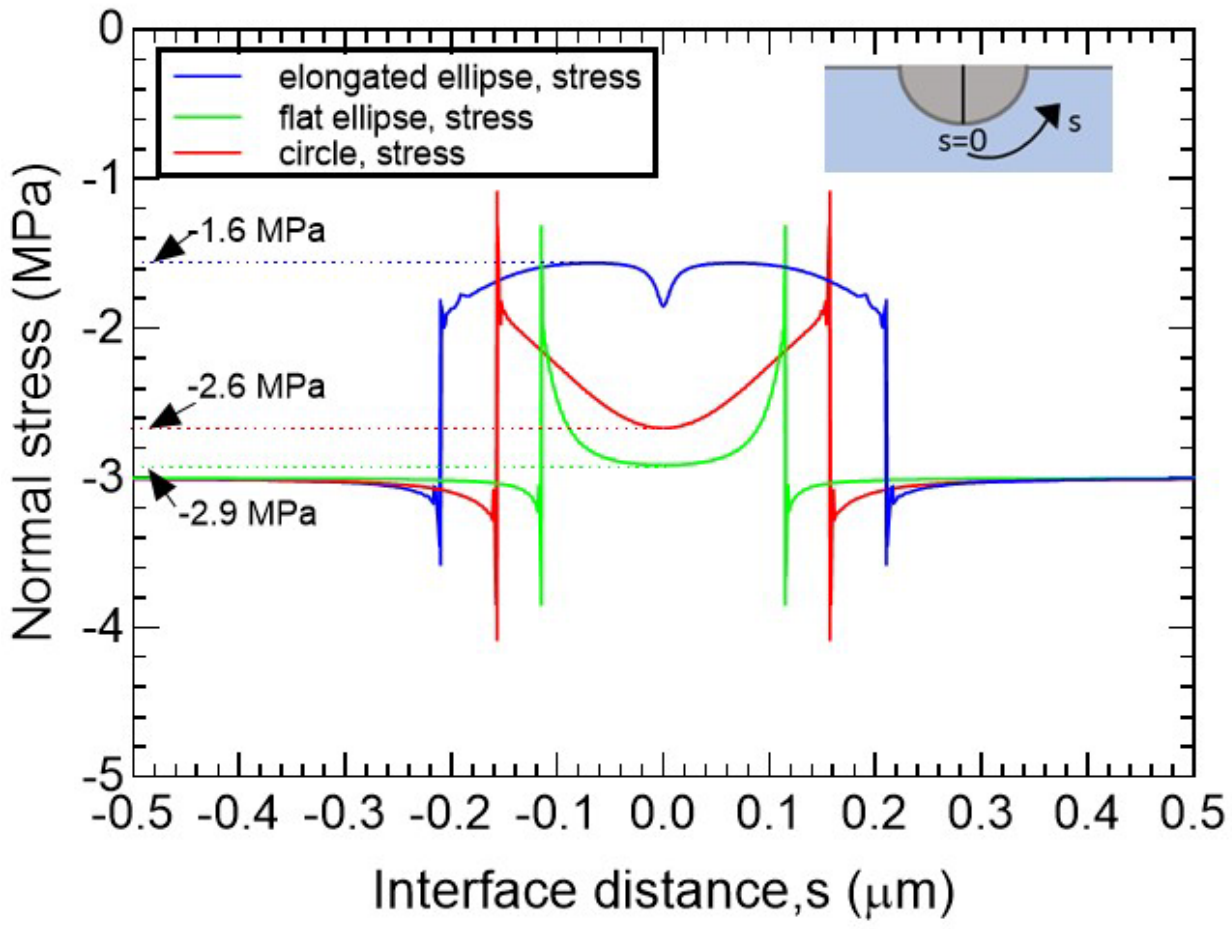

**Fig. 4** Normal stress distribution around the roughness for elongated ellipse, flat ellipse and circle.

The reported 3 MPa corresponds to the normal compressive stress component rather than the von Mises equivalent stress. Although this value exceeds the yield strength of Li, yielding is governed by the deviatoric stress state, as described by the von Mises yield criterion. In the present simulations, the equivalent stress remains close to the yield limit, while the hydrostatic stress component can exceed the yield strength without inducing additional plastic deformation. The coupled equations of mechanical equilibrium, the yield criterion, and the plastic flow rule are solved simultaneously. Furthermore, the lithium metal is mechanically constrained by the surrounding solid electrolyte, which restricts plastic flow and permits elevated compressive stresses to persist under equilibrium conditions. Although plastic deformation occurs within the lithium bulk, it does not completely homogenize the interfacial stress distribution because of the geometric roughness and mechanical boundary constraints.

### *3.2 Interfacial Normal Stress Impact on Current Distribution*

#### *3.2.1 Plating Condition*

Given the boundary conditions in Figure 2(a), the governing equation is solved for the case of plating. In Figure 5, the overpotential distribution along the interface for various geometries of roughness is presented. For a smooth flat Li/SE interface with no defects, the overpotential is -25.918 mV estimated using $i_{BV}$ =-100μA/cm$^2$ in Eq. (10) without mechanical effects. When a flat elliptical roughness is introduced at the interface, the overpotential increases slightly to -25.881mV (Fig. 5b). For an elongated elliptical roughness, the overpotential rises further to -25.516 mV (Fig. 5c). The circular roughness produces an overpotential that lies between these two cases (Fig. 5a). The variation in overpotential among the different roughness geometries partly arises from differences in the total interfacial area. For the flat ellipse, the interfacial area remains close to that of a smooth flat, whereas the elongated ellipse increases the interfacial area and consequently reduces the magnitude of the interfacial current density, $i_{BV}$ .

As shown in Figure 5a, both mechanical stress and the shape of the roughness influence overpotential. By enlarging the region around the roughness, as in Figures 5(b) and 5(c), the local variations in the overpotential distribution become visible. However, these variations remain small.

For a flat smooth Li/SE interface with no defects, the interfacial normal stress is 3 MPa compression. The impact of mechanical stress on the overpotential expressed as $\exp\left(-\frac{\Omega_{etrode}^{Li}\sigma_n}{2RT}\right)$ in Eq. (10) becomes approximately -0.2 mV. Regardless of the extrusion geometry, the interfacial normal compressive stress remains 3 MPa away from the roughness. This results in the fact that when mechanical effects are included, the magnitude of the overpotential increases by 0.2 mV in all cases as shown in Figure 5(a). As discussed previously, the interfacial area for the elongated ellipse is about 1% larger than that for the flat ellipse. As a result, the average interfacial normal current density for the elongated ellipse becomes roughly 1% lower. A 1% decrease in normal current density corresponds to an overpotential change of approximately 0.26 mV, which is similar in magnitude to the mechanical effect (~0.2 mV). Due to the non-uniform current density distribution along the rough surface (shown later in Fig. 6), the numerical results indicate that the elongated ellipse produces an overpotential that differs from the flat ellipse by about 0.4 mV.

These results are not intended to assess the direct impact of overpotential variations on battery durability or performance. Rather, the objective of the present study is to identify the dominant mechanism responsible for non-uniform lithium deposition. The simulations suggest that mechanical stress is the primary contributor to deposition non-uniformity (shown later in Fig.6).

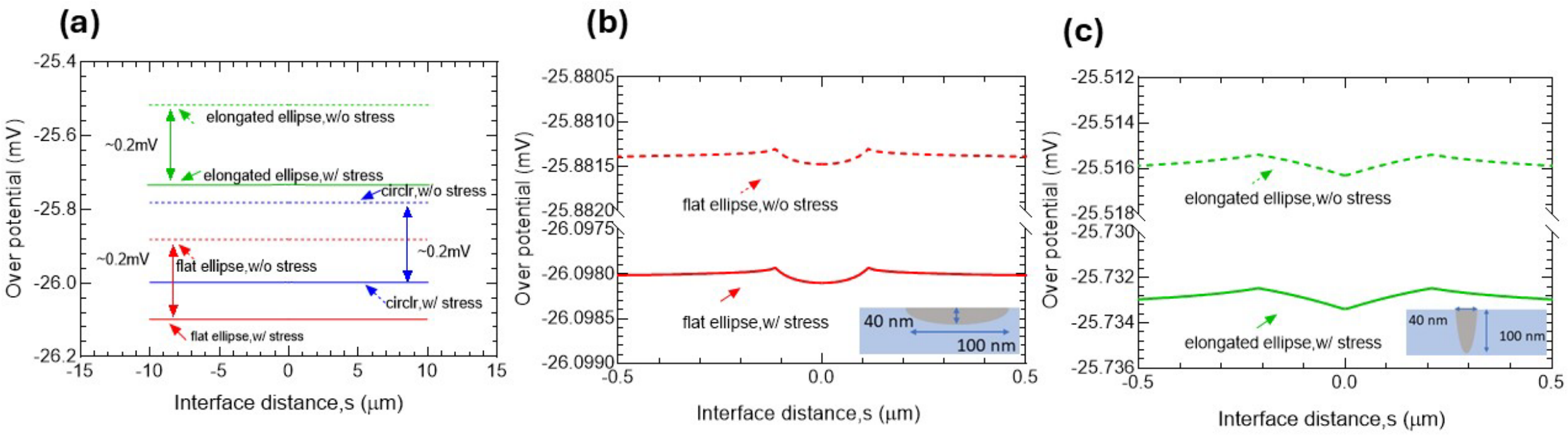

**Fig. 5 (a)** Interface overpotential distribution along the roughness for different geometries under plating condition; **(b)** Focused overpotential distribution along the interface of the flat ellipse; **(c)** Focused overpotential distribution along the interface of the elongated ellipse.

Figure 6 illustrates the magnitude of normal current density and normal stress distributions across different roughness geometries, both with and without the influence of mechanical stress. When mechanical effects are not considered, there is minimal change in the normal current density indicating almost uniform Li-metal deposition. However, when mechanical effects are taken into account, a noticeable change in the normal current distribution is observed. This variation corresponds to non-uniform Li-metal deposition. If mechanical stress is neglected, Li ion deposition rate remains almost uniform even on a rough surface.

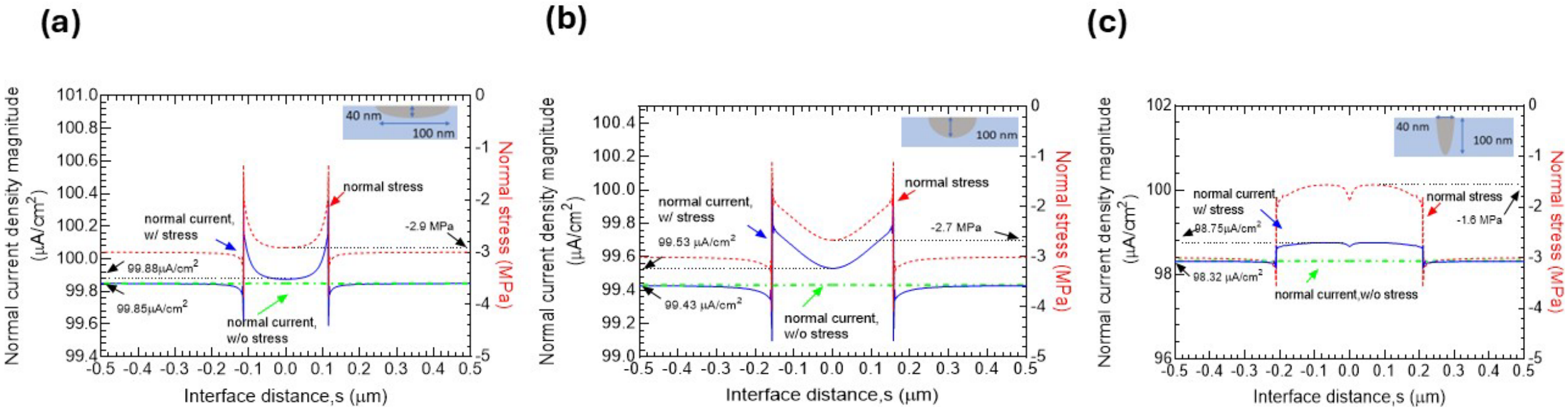

**Fig. 6** Magnitude of normal current and normal stress distributions under the plating condition along the interface for (a) flat ellipse, (b) circle, (c) elongated ellipse.

The magnitude of normal current density at the center of the flat elliptical roughness (~99.88 μA/cm²) is nearly identical to that observed in the region away from the roughness (~99.85 μA/cm²). The difference in the central normal current density between the stressed and unstressed cases is minimal, approximately 0.03 μA/cm² that is, 0.03% for the flat elliptical roughness and 0.1 % for the circular roughness. Both the flat elliptical and circular geometries exhibit locally elevated normal current densities near the edges of the roughness features. In contrast, the magnitude of normal current density at the center of the elongated elliptical roughness (~98.75 μA/cm²) exceeds that in the region away from the roughness (~98.32 μA/cm²), indicating a higher degree of current concentration around the elongated ellipse. For this geometry, the difference between the stressed and unstressed cases increases to approximately 0.5 %. Moreover, while the flat and circular roughnesses show high current densities concentrated at their edges, the elongated ellipse exhibits elevated current densities along its contour.

When comparing the values of the normal current density magnitude away from the various roughness shape, the flat elliptical case (99.85 μA/cm$^2$) is the closest to the value without any roughness (100 μA/cm$^2$) since the interfacial area does not increase very much due to the roughness, whereas the elongated elliptical case (~98.32 μA/cm$^2$) is the lowest since the interfacial area increase most. And for all roughness geometries, the plating current density distribution around surface roughness exhibits a spatial pattern similar to that of the interfacial normal stress distribution.

To provide a simple physical interpretation, the local deposition rate is proportional to current density via Faraday's law. Therefore, a relative variation in current density (e.g., ~0.1–0.5%) directly translates to a comparable relative difference in local deposition rate. Over extended cycling, such persistent differences can accumulate and lead to measurable height variations. However, capturing this long-term evolution would require coupling the present model to moving-boundary or growth simulations, which is beyond the scope of the current study.

The current distribution in the SE for different surface roughness shapes is shown in Figure 7. The small roughness features of various shapes disturb the current density at the interface, as indicated by the flow of current represented by red arrows. The lengths of the arrows correspond to the magnitude of the current density, which increases near the edges of the protrusions. The embedded magnified images highlight the local variation in current density along each roughness feature. For the elongated elliptical shape, the variation in current density is more pronounced (Fig. 7c), as indicated by the noticeable change in arrow length. In contrast, the flat elliptical shape (Fig. 7a) shows the least change in current density magnitude. These results show that the elongated elliptical shape tends to enhance Li deposition around the extrusion while the Li deposition remains almost uniform around the flat ellipse except the edges.

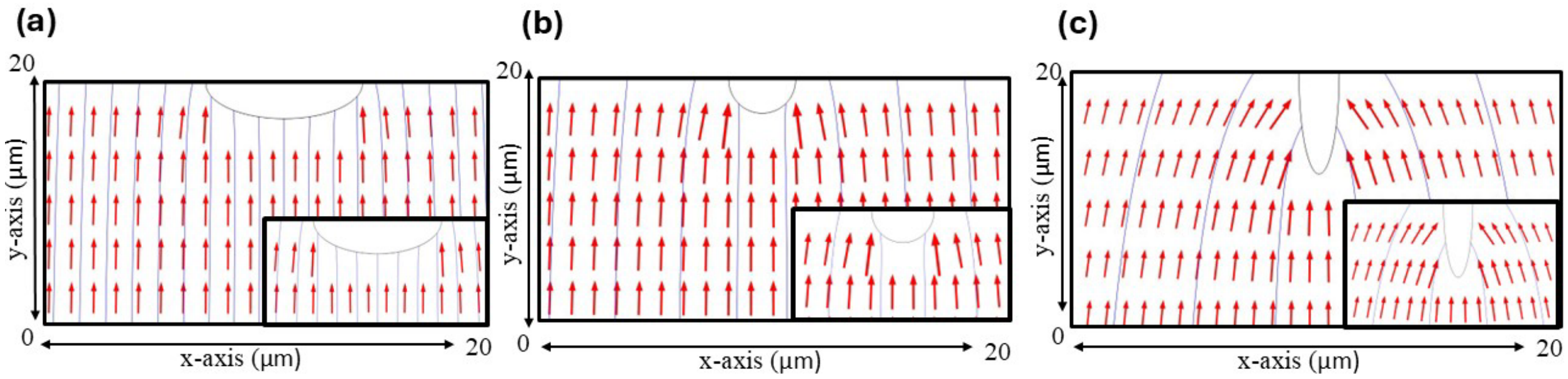

**Fig. 7** Current distribution during plating in solid electrolyte around **(a)** flat elliptical, **(b)** circular, **(c)** elongated elliptical roughness. The inset figures show closer views of current distributions near the roughness. The arrowhead indicates current direction.

### *3.2.2 Stripping Condition*

The overpotential distribution along the nonuniform Li/SE interface under stripping conditions has also been examined. The effects of both mechanical stress and surface roughness on the overpotential are clearly evident when comparing the stressed and unstressed cases, together with their corresponding interfacial surface areas. Similar to the plating condition, surface roughness and mechanical stress exert comparable influences on the interfacial kinetics (Supplementary Information, S3). For a smooth flat Li/SE interface with no roughness, the overpotential is 25.917 mV estimated using $i_{BV}$ =100 μA/cm² as described in the previous section without mechanical effects. Introducing a flat elliptical roughness reduces the overpotential slightly to 25.881 mV, while an elongated elliptical roughness further reduces it to 25.516 mV. The circular roughness yields an overpotential between these two values. These differences are partly governed by the total interfacial area associated with each geometry. The flat ellipse produces an area similar to the smooth interface, whereas the elongated ellipse increases the interfacial area and correspondingly lowers the interfacial current density. As in plating, the local variations in the overpotential distribution around the roughness remain small (Fig. S3 (b), (c)).
For the smooth flat Li/SE interface, the interfacial normal stress is 3 MPa in compression. When mechanical effects are included, the resulting change in overpotential for the flat interface is ~0.6 mV. As in plating, this leads to a uniform decrease of 0.6 mV across all three cases of surface roughness (Fig. S3).
To gain deeper insight into the stripping behavior, further computational modeling was performed using an elongated elliptical surface roughness. Figure 8(a) shows the distributions of normal current density and normal stress across the protrusion, while Figure 8(b) illustrates the direction and magnitude of the current density vectors. A lower magnitude of normal current density is observed along the contour of the elongated protrusion (~97.17 μA/cm²) compared to the case without mechanical stress (~98.33 μA/cm²), indicating slower Li stripping in the stressed condition. Combined with the prior analysis of high Li deposition rate around the elongated ellipse, growth of elongated ellipse is anticipated.

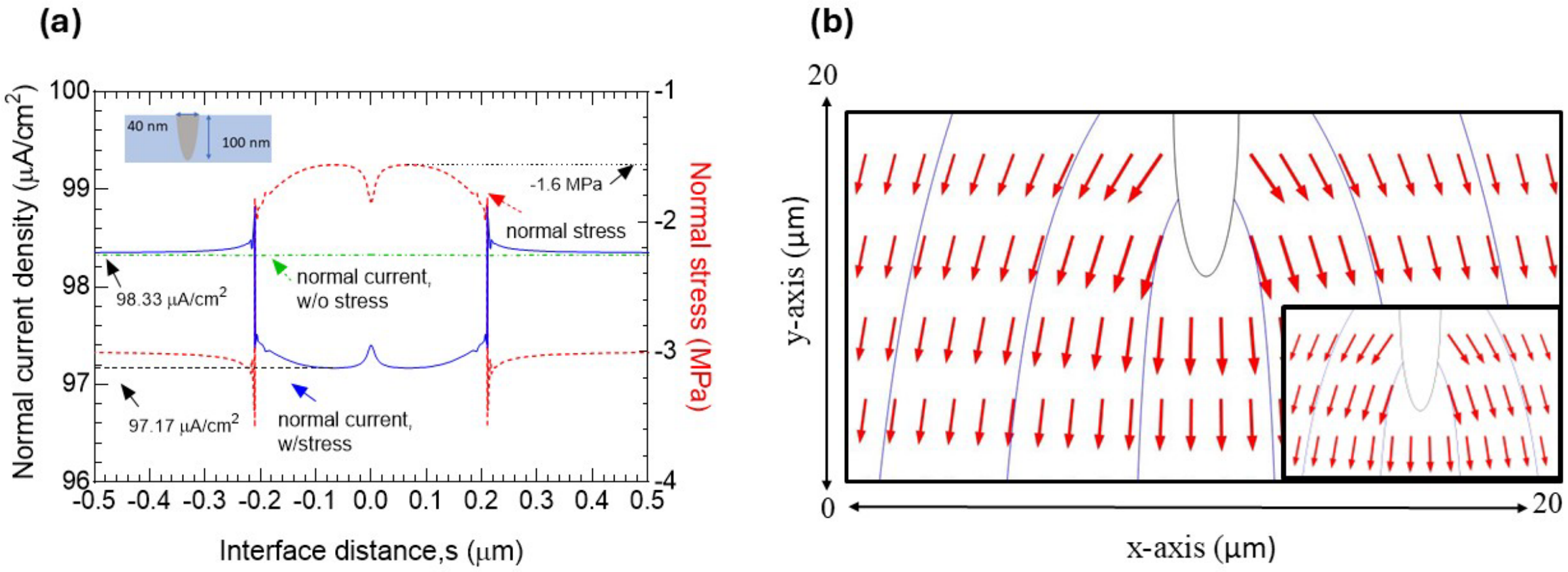

**Fig. 8 (a)** Normal current and normal stress distributions along the interface under the stripping condition; **(b)** Current distribution in the whole SE during stripping. The inset figure in (b) shows a closer view of current distributions near the roughness. The arrowhead indicates current direction.

### *3.3 Influence of Exchange Current and Applied Current Density Variations*

### *3.3.1 Plating Condition*

Numerous studies on metal/electrolyte interface in ASSBs have primarily focused on reducing interfacial contact resistance. Higher exchange current density corresponds to a lower interfacial resistance. However, experimentally reported exchange current densities are subject to uncertainties. To systematically assess the influence of this parameter on electrodeposition behavior, we performed additional simulations over a range of exchange current density values while keeping the applied current density at 100 μA/cm² (Supplementary Information S4) and then fixed the exchange current density at the experimentally obtained value while increasing the applied current density by a factor of 100. (Supplementary Information S5).

When the exchange current density is reduced to 0.01 times its experimental value, the interfacial resistance increases substantially. If smooth flat surface is assumed, the overpotential becomes -239.563 mV using $i_{BV}$ = -100 μA/cm$^2$. The overpotential difference between the flat and elongated elliptical geometries increases from approximately 0.4 mV (Fig. 5a) to approximately 0.8 mV (Fig. S4(a)). Figure S4(a) indicates that the effect of stress remains negligible in this regime.

In contrast, when the exchange current density is very large (100× the experimental value), the interfacial resistance is small. In such case, linearized Butler-Volmer equation can be used and is further discussed in section 3.4. In this regime, the kinetics effect on overpotential becomes negligible relative to the effect of mechanical stress. A small decrease in the magnitude of normal current density corresponds to an overpotential change represented by the first term on the right-hand side of Eq. (15). This term becomes insignificant compared to the ~0.4 mV contribution from mechanical stress.

At very low exchange current densities (Supplementary Information S4), the spatial distribution of normal current density with and without mechanical stress remains nearly identical (Fig. S5(a)) and does not exhibit patterns similar to stress distribution, indicating that mechanical stress has a negligible effect on the Li deposition rate. However, at very high exchange current densities (100× the experimental value), the magnitude of the normal current density near the elongated elliptical interface increases markedly from ~98.2 μA/cm² (without stress) to ~167.7 μA/cm² (with stress), demonstrating the strong influence of mechanical stress on Li deposition kinetics. The Li deposition rate (i.e., normal current density magnitude) is elevated over the entire surface for the elongated elliptical geometry (Fig. S5(b)).

Further investigations were conducted by increasing the applied current density by a factor of 100 while keeping the exchange current density fixed at the experimentally obtained value of 95 μA/cm² to examine its impact on Li deposition (see Supplementary Information S5). The resulting overpotential distributions for various roughness geometries, with and without mechanical effects, nearly overlap, indicating that the influence of stress remains negligible (Fig. S6(a)). The overpotential difference between flat elliptical and elongated roughness geometries is approximately 0.9 mV (Fig. S6(a)), and the spatial distribution of normal current density does not follow the normal stress pattern (Fig. S6(b)).

### *3.3.2 Stripping Condition*

The influence of exchange current density was also examined for values of 0.01× and 100× the experimental exchange current density under stripping conditions, both of which exhibited similar qualitative trends (Supplementary Information S6) and then fixed the exchange current density at the experimentally obtained value while increasing the applied current density by a factor of 100. (Supplementary Information S7).

When the exchange current density is very small (0.01× the experimental value), the interfacial resistance increases. Under these conditions, the contribution of mechanical stress to the overpotential becomes negligible as shown in Figure S7(a). Assuming a smooth flat surface, Eq. (13) predicts an overpotential of 239.563 mV using $i_{BV}$ = 100 μA/cm$^2$. In contrast, at very large exchange current densities (100× the experimental value), the interfacial resistance is minimal, so linearization of Butler-Volmer equation is applicable here. In this regime, the influence of surface geometry on overpotential becomes negligible compared to mechanical stress, which contributes to overpotential change of approximately 0.4 mV.

At very low exchange current densities (Fig. S8a), the spatial distribution of normal current density with and without mechanical stress remains nearly identical, indicating that mechanical stress has a negligible effect on the Li stripping rate. However, at very high exchange current densities (100× the experimental value), the interfacial resistance is lower. The magnitude of the normal current density near the elongated elliptical interface decreases markedly from ~97.88 μA/cm² (without stress) to ~27.51 μA/cm² (with stress), demonstrating the influence of mechanical stress on Li stripping (Fig. S8b). The magnitude of the stripping current density around the elongated ellipse becomes very low. Once deposited, the Li is not stripped significantly, which in turn leads to very rapid growth.

Additional simulations were performed by increasing the applied current density by 100 times while keeping the exchange current density fixed at the experimental value of 95 μA/cm² for the stripping case (Supplementary Information S7). The resulting overpotential distributions for different surface roughness geometries, with and without mechanical effects, show no significant influence of stress compared with kinetic effects (Fig. S9(a)). The spatial distribution of normal current density closely follows the normal stress pattern (Fig. S9(b)), but the impact of mechanical stress on the normal current density remains negligible.

### *3.4 Justification for the Linearized Butler-Volmer Formulation*

While all the numerical results presented here are based on the nonlinear form of the Butler-Volmer equation (10), the linearized form of Eq. (14) and (15) will be further discussed in this section. A comparison of the nonlinear and linearized version of the Butler-Volmer equations is shown in Figure 9(a) using the experimentally measured exchange current density. When the interfacial current density $(i_{BV})$ is zero, the mechanics modified overpotential $(\tilde{\eta}')$ becomes zero, resulting in the mechanics-free overpotential $(\tilde{\eta})$ becoming $\frac{\Omega_{etrode}^{Li}\sigma_n}{F} = -0.404$ mV if $\sigma_n = -3$ MPa as shown in Figure 9(b). If the compressive stress is 10 MPa, this shift of $\tilde{\eta}$ increases to –1.347 mV.

As demonstrated in the parametric studies presented in section 3.3, when the exchange current density becomes extremely large, the stress term gains significance, and the magnitudes of (mechanics-free) overpotential becomes small. In such cases, the use of the linearized form becomes feasible. In the linearized equation (15), interfacial kinetics term $(R_{\text{int}} i_{BV})$, and mechanical stress term $\left(\frac{\Omega_{etrode}^{Li}\sigma_n}{F}\right)$, are expressed as a simple addition. This straightforward expression highlights the relative significance of mechanical stress in a simplified form. Assuming that the contribution from stress is -0.404 mV with a compression of 3 MPa, if the magnitude of the interfacial kinetics term becomes 1mV, then the stress contribution becomes significant. This leads to the range of $\left(\frac{i_{BV}}{i_o}\right)$ is 0.0389 or less. Even when this ratio is close to 1 using the experimental value of the exchange current density, the error resulting from linearization is minimal as illustrated in Figure 9. It is also shown in section 3.3, when the exchange current density becomes large, then the normal current density variation around the rough surface becomes large. The high ion deposition rate and low ion stripping rate are observed around the elongated elliptical roughness interface. This nonuniform ion deposition/stripping rate enhances the growth of the dendrite. The key reason for this is attributed to the interfacial normal stress variation around the surface roughness.

Reducing interfacial resistance is crucial for optimizing battery cycle performance. When interfacial resistance decreases, the ratio of $\left(\frac{i_{BV}}{i_o}\right)$ becomes smaller, leading to significant stress contribution. However, the nonuniform distribution of interfacial normal stress results in nonuniform ion deposition and stripping rates, which are undesirable and contribute to dendrite growth. To address this trade-off, a protective layer can be used as a buffer zone. The linearized version of the Butler-Volmer equation can be employed for a simplified analysis of these cases. While measuring $i_{BV}$ distribution along the interface with roughness is not possible, $i_{app}$ is known. By estimating the order of magnitude of $i_{BV}$ from $i_{app}$, insights into the interface's behavior can be gained.

Battery operation under high current density is important. As shown in Section 3.3 and Figure S5 and S8, the stress effect can be ignored in this case. However, if there is a significant improvement in increasing the exchange current density, the stress effect can become important.

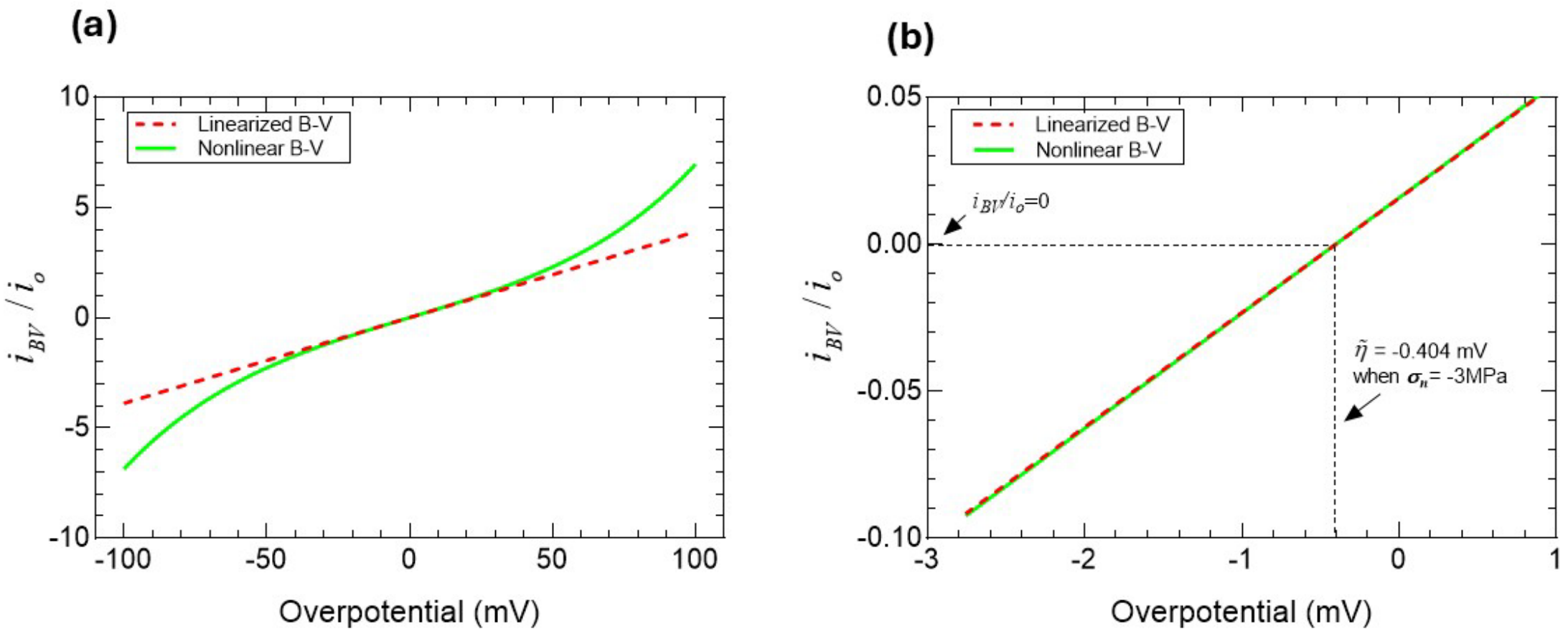

**Fig. 9** **(a)** Comparison of the nonlinear and linearized stress-induced Butler-Volmer relations, and **(b)** enlarged view of the overpotential region near zero applied current density.

### *3.5 Protective Coating*

To investigate plating/stripping in the presence of a protective layer shown in Figure 2(d) and Figure 2(e), a similar analysis described in section 3.1 to section 3.2 was conducted and the results are compared with those without a protective coating.

As illustrated in Figure 10(a), the application of a protective lithium nitride ($Li_3N$) layer further reduces the magnitude of the local normal compressive stress from 1.6 MPa (without the layer) to 0.6 MPa (with layer) around the elongated elliptical surface roughness. Concurrently, the magnitude of the normal current density during lithium plating increases approximately to 99.44 μA/cm² from 98.75 μA/cm² in the presence of the $Li_3N$ layer. On the other hand, a distinct change in current distribution is evident while stripping, with the normal current density magnitude decreases to approximately 95.93 μA/cm² from 97.17 μA/cm² in the presence of the layer around the rough interface, as shown in Figure 10(b).

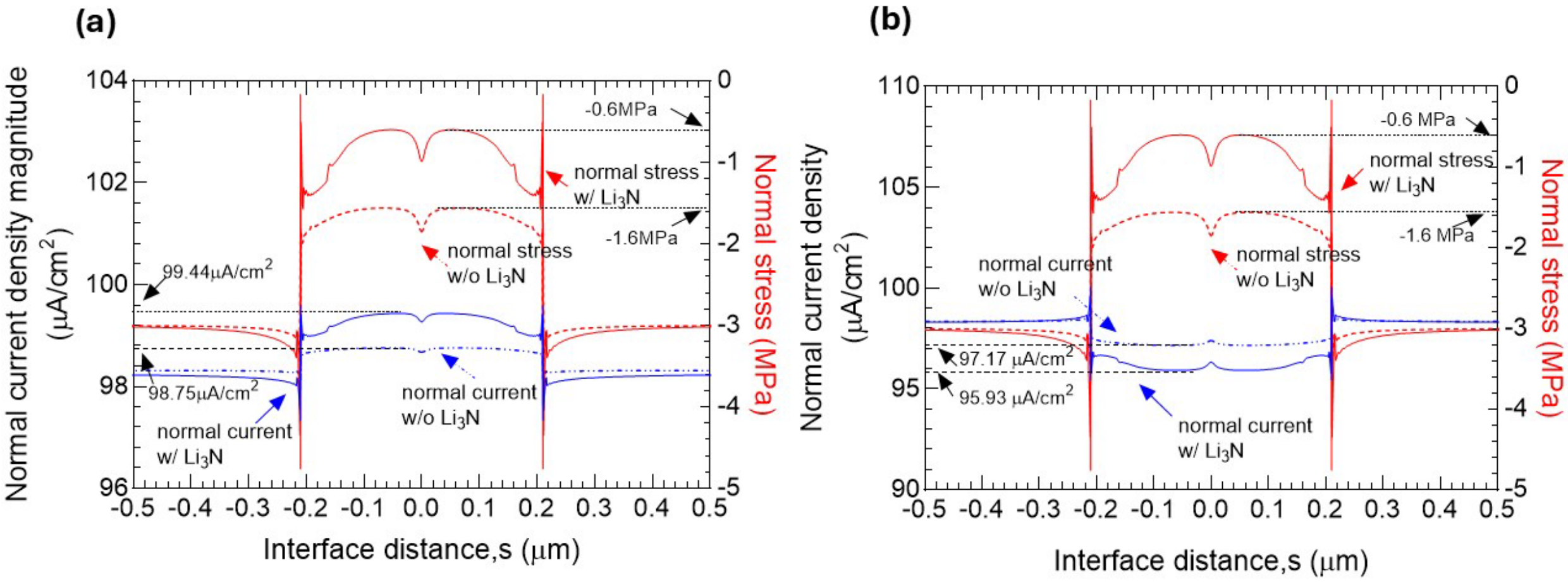

**Fig. 10** Effect of the $Li_3N$ interlayer on normal current density magnitude and normal stress distributions along the interface of an elongated elliptical roughness considering mechanical effect under **(a)** plating and **(b)** stripping conditions.

It is well established that protective coatings play a critical role in suppressing Li dendrite growth and improving overall battery performance by acting as buffer layers that isolate SEs from the Li metal anode [66]. The effectiveness of such coatings, however, strongly depends on interfacial compatibility and geometric factors, particularly the thickness ratio between the protective layer and the SE. For example, preliminary experimental studies employing a thick SE (~700 μm) together with a thin $Li_3N$ interlayer (~3 μm) demonstrated good cyclic performance [55, 67] whereas numerical simulations using a much thinner SE (20 μm) exhibited a significant reduction in effective ionic conductivity. As shown in supplemental information, the effective ionic conductivity for the SE with the protective layer is highly sensitive to the thickness ratio between the $Li_3N$ protective layer (t) and the argyrodite SE (L). When t = 3 μm and L = 20 μm, a notable degradation is observed: $\sigma_{eff}^{+} = 0.33898\sigma_{SE}^{+}$, (Supplementary Information S8).

Under the assumption of a perfectly bonded interface, a softer interfacial coating material can be advantageous because its elastic properties more closely match those of lithium metal, thereby reducing the variation in normal stress distribution around surface roughness. In contrast, the protective layer $Li_3N$ considered in this study is comparatively harder, resulting in considerable normal stress variation, as shown in Figure 10. Softer interfacial materials may promote a more uniform ion plating/stripping rate and suppress dendrite growth, even when the interfacial resistance remains relatively low. In line with this concept, recent work by Yao et al. [68] demonstrated that use of Ag-C composite which is a softer interlayer improves interfacial mechanical contact, promotes uniform current distribution, and effectively suppresses Li dendrite growth.

## 4.Conclusion

In this study, a continuum theory-based 2D model was developed to investigate Li deposition, emphasizing the coupled effects of surface roughness and mechanical stress. The results highlight the strong interplay between electrochemical response and mechanical response at the Li/SE interface, providing insights into electrodeposition/stripping behaviors based on local current density distributions without time dependence.

The present analysis focuses on the initial stages of deposition. Therefore, the results should be interpreted as indicating a mechanistic tendency toward nonuniform deposition rather than a quantitative prediction of growth over cycling. The key findings are summarized as follows.

(**1**) Under all the conditions investigated, ignoring the mechanical effect, the current density concentration around the roughness is not significant during the early stage of dendrite development, indicating uniform Li plating without dendrite formation, which is in contradiction to experimental findings [69,70]. Therefore, including the mechanical contribution in the model and coupling it with electrochemistry are indispensable.

(**2**) However, this does not imply that stress is always significant. It becomes dominant only when the ratio of the interfacial current density normalized by the exchange current density is significantly small (0.0389 or less for a compression stress of 3 MPa). As demonstrated in Section 3.2, even if this ratio approaches close to 1, the stress term still plays a significant role and cannot be ignored. As interfacial resistance increases with battery aging, the exchange current density decreases, and the stress effects on the interfacial kinetics diminish.

(**3**) The geometry of surface roughness governs the distribution of interfacial normal stress, which in turn strongly influences local deposition behavior. Even though no time dependence is considered, surface roughness with elongated shape tends to grow based on the interfacial deposition/stripping current density.

(**4**) Reducing interfacial resistance is crucial for long-term battery performance. When the interfacial resistance is low, the exchange current density increases, and the stress term becomes important. However, in this case, nonuniform stress distribution leads to nonuniform ion deposition/stripping rates around the surface roughness, resulting in dendrite growth.

(**5**) To mitigate this issue, a soft buffer layer may be used to reduce the nonuniformity of normal interfacial stress distribution while maintaining reduced interfacial resistance.

**Acknowledgement**

K.I. acknowledges the financial support of NSF National Research Traineeship Program, The Ohio State EmPOWERment Program (Grant # 1922666). This work was also supported in part by The Ohio State University Materials Research Seed Grant Program, funded by the Center for Emergent Materials, an NSF-MRSEC, grant DMR-2011876, the Center for Exploration of Novel Complex Materials, and the Institute for Materials and Materials Research. The authors also acknowledge Smart Vehicle Concepts Center (www.SmartVehicleCenter.org), a graduated National Science Foundation Industry–University Cooperative Research Center initially established under Grant NSF IIP 1738723.

## Supplementary Information

### Effect of Stress and Surface Roughness on Electrodeposition in All-Solid-State Batteries: A Computational Investigation

**Kaniza Islam[1], Ayush Morchhale[1] Jung-Hyun Kim[1], Yanzhou Ji[2,†] , Noriko Katsube[1]**

[1] *Department of Mechanical and Aerospace Engineering, The Ohio State University, Columbus, OH, 43210, United States*

[2] *Department of Materials Science and Engineering, The Ohio State University, Columbus, OH, 43210, United States*

[†] Email: ji.730@osu.edu

## S1. Mesh Convergence and Numerical Implementation

All simulations were performed in COMSOL Multiphysics using a two-dimensional finite element formulation with quadratic elements and a free triangular mesh. To accurately resolve the stress and electrochemical fields near the surface roughness, default local mesh refinement was applied in the vicinity of the roughness features at the Li/SE interface. The governing equations were solved using a stationary solver with a relative tolerance of $10^{-3}$.

A mesh-convergence study was conducted for the elongated elliptical roughness geometry to verify that the reported results are independent of the numerical discretization (Figure S1).

It should be noted that geometries containing sharp corners can produce stress singularities in continuum finite element models, leading to localized stress amplification near the corner points. However, the conclusions of the present study do not rely on these localized singular features. The reported trends and interpretations are based on the overall distributions of stress, overpotential, and current density along the interface rather than the peak values at corner singularities. Consequently, the key findings remain unaffected by the localized numerical stress concentrations associated with sharp geometric features.

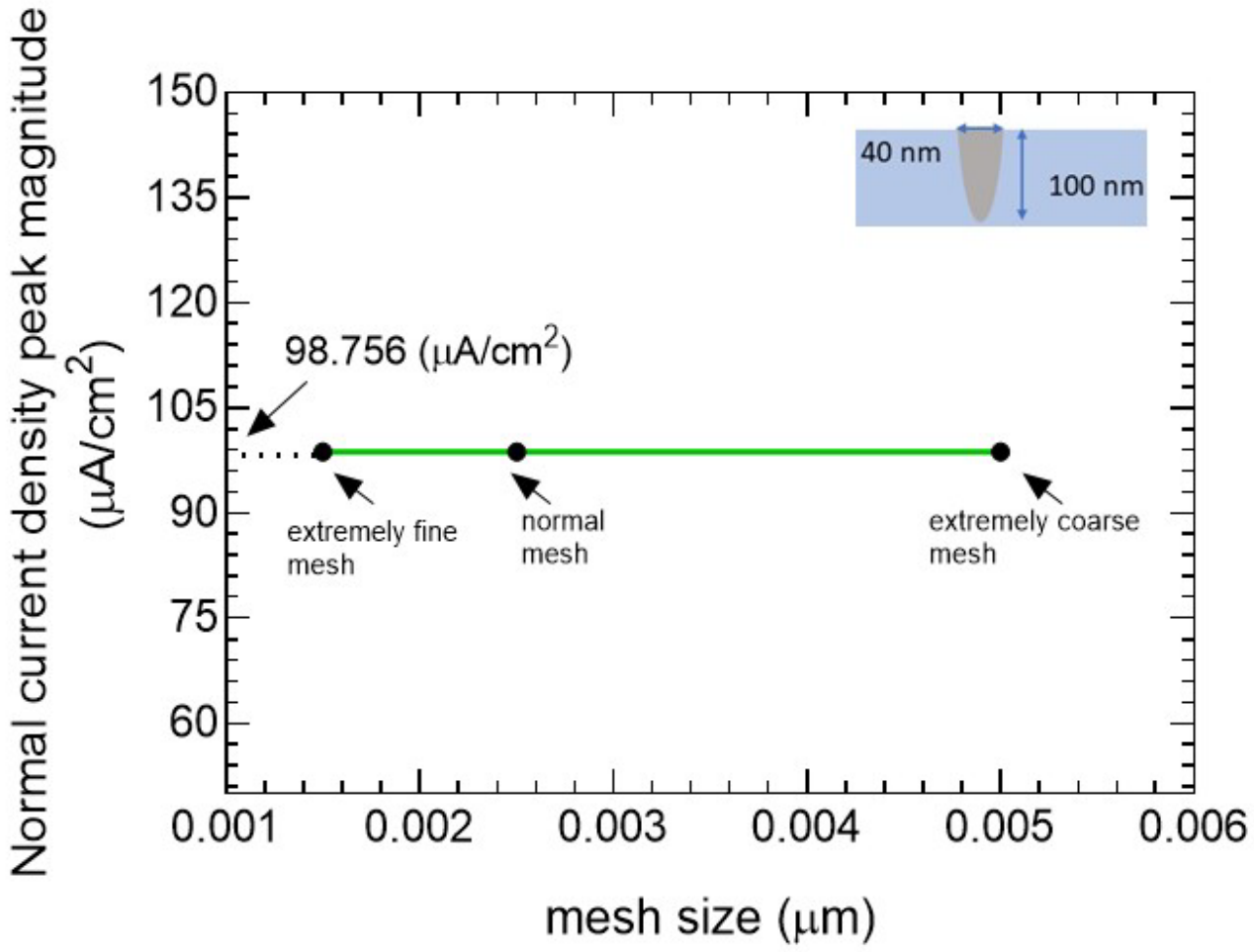

**Fig. S1** Mesh refinement study demonstrating convergence of the peak normal current density near the roughness.

## S2. Determination of Exchange Current Density

The Tafel plots shown in Figure S1 were used to compare the electrochemical kinetics of Li and $Li_3N$–Li symmetrical cells. The exchange current density ($i_0$) was determined by extrapolating the linear region of each Tafel plot to zero overpotential. The $Li_3N$–Li symmetrical cell exhibits a smaller Tafel slope and a higher exchange current density (133 ± 9 μA cm$^{-2}$) than the Li symmetrical cell (95 ± 6 μA cm$^{-2}$). This result indicates that the electrochemically stable Li–$Li_3N$/SE interface facilitates more efficient charge-transfer kinetics during lithium plating and stripping.

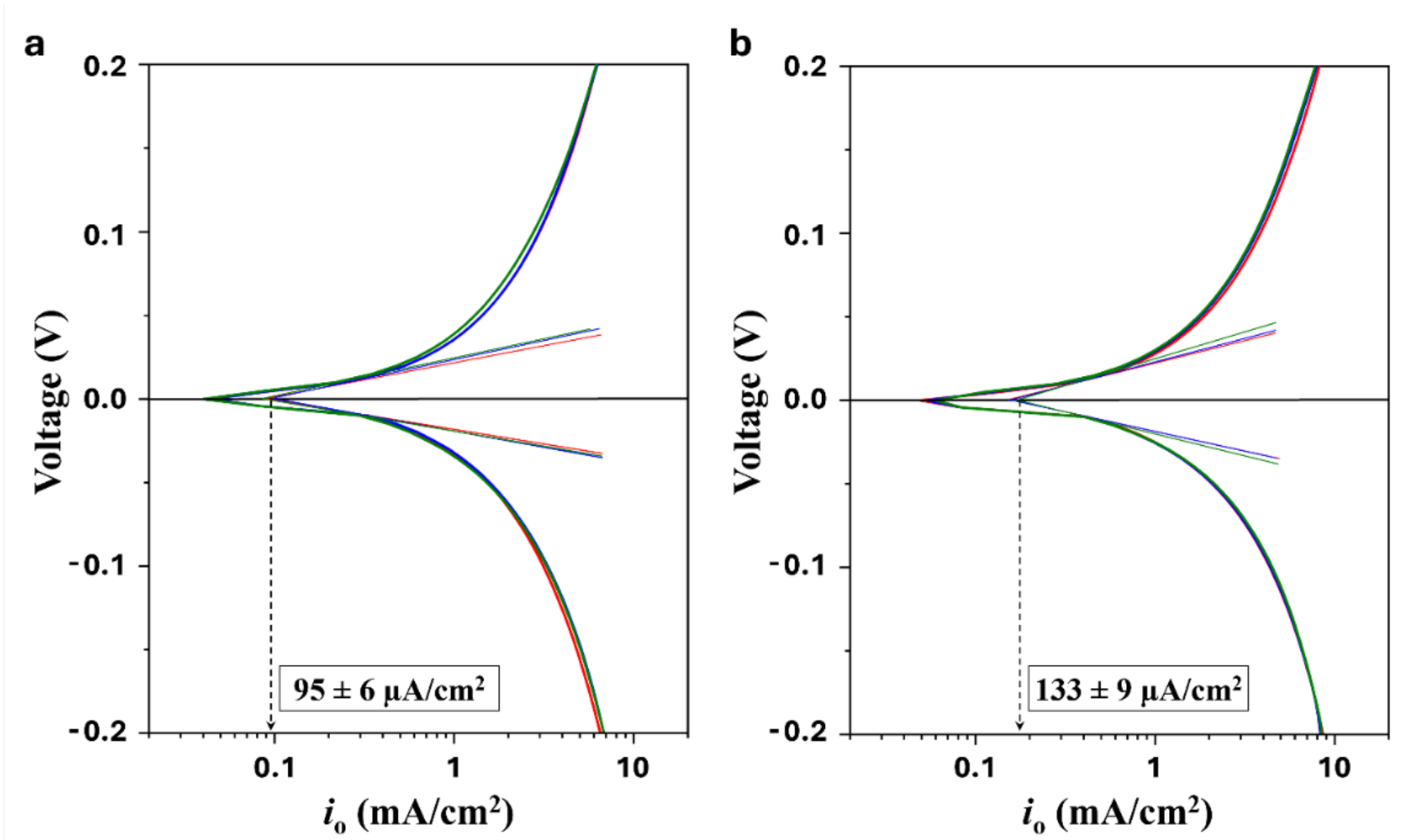

**Fig. S2** Tafel plots obtained from (a) Li and (b) $Li_3N$-Li symmetric cells at room-temperature. Data from three individual cells are plotted using different colors.

## S3. Interfacial Normal Stress Impact, Stripping

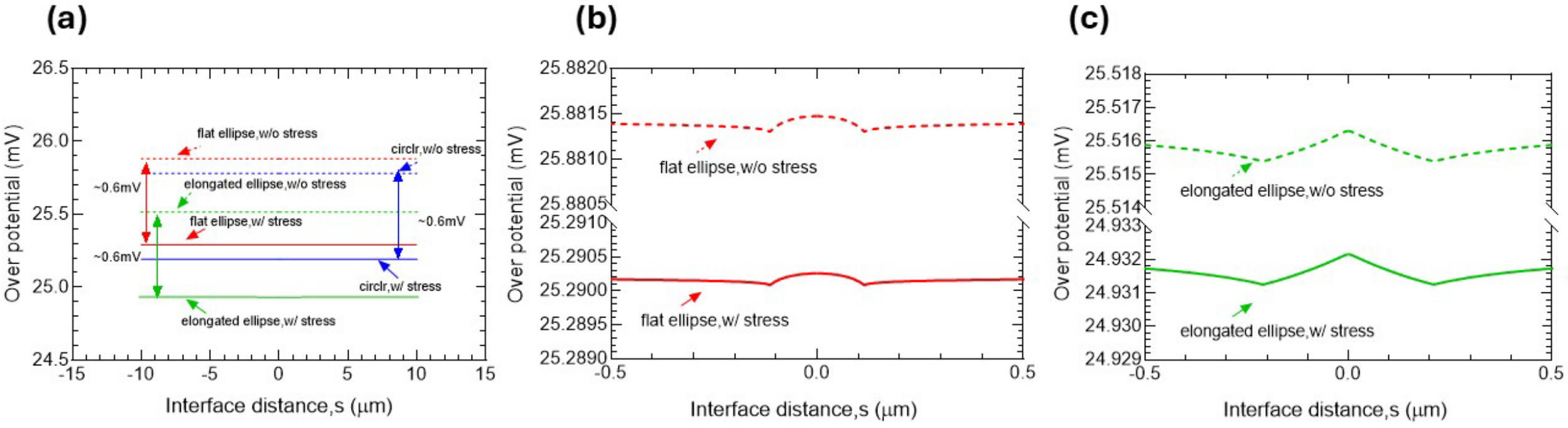

**Fig. S3 (a)** Interface over potential distribution along the roughness for different geometries for stripping condition; **(b)** Focused over potential distribution along the interface for flat ellipse; **(c)** Focused over potential distribution along the interface for elongated ellipse.

## S4. Influence of Exchange Current Density Variations, Plating

The current density associated with plating and stripping at the interface is governed by the surface overpotential, defined as the difference in electric potential between SE and Li via the modified Butler-Volmer equation.

Under plating conditions, for a smooth flat interface with an exchange current density of $i_0$ = 0.95 μA/cm$^2$ (0.01× the experimental exchange current density) the overpotential is -239.563 mV, with mechanical stress when $i_{BV}$ is -100 μA/cm$^2$, as given by Eq. (13). When a flat roughness is introduced, the overpotential value remains close to that of a smooth flat interface. And when the interfacial geometry becomes an elongated ellipse, the average normal current density decreases by about 1% compared to a flat surface. This results in a slightly higher overpotential of approximately -238.33 mV, as shown in Fig. S1(a). Regardless of the roughness geometry, the effect of mechanical stress on interfacial kinetics remains negligible in this regime.

In contrast, when the exchange current density is increased to $i_o$ = 9500 μA/cm² (100× the experimental value), the inclusion of mechanical stress leads to a measurable overpotential change of approximately 0.4 mV using the parameters listed in Table I. This stress-induced increase in the magnitude of overpotential is observed for all geometries, as shown in Fig. S3(b).

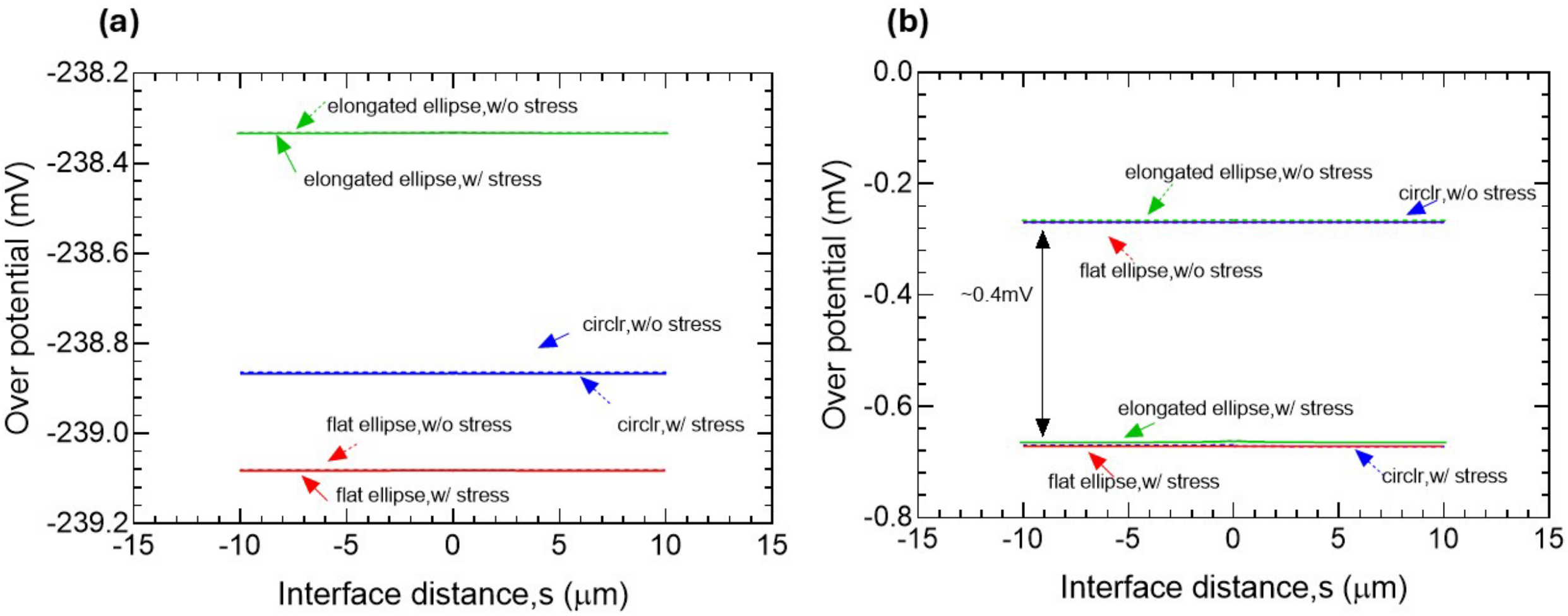

**Fig. S4** Interface over potential distribution along the roughness for different geometries **(a)** 0.01× the experimental exchange current density; **(b)** 100× the experimental exchange current density for plating condition.

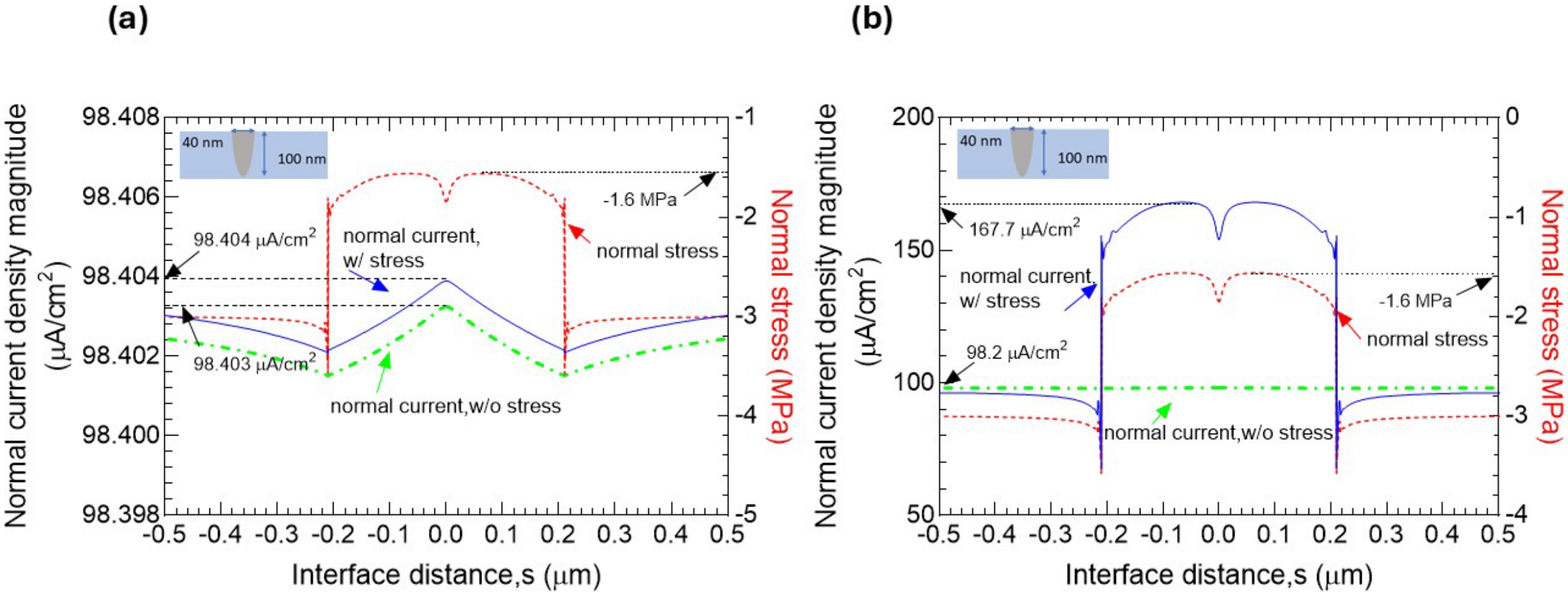

**Fig. S5** Magnitude of normal current and normal stress distribution along the Interface for **(a)** 0.01× the experimental exchange current density, **(b)** 100× the experimental exchange current density for plating condition.

## S5. Influence of Applied Current Density Variations, Plating

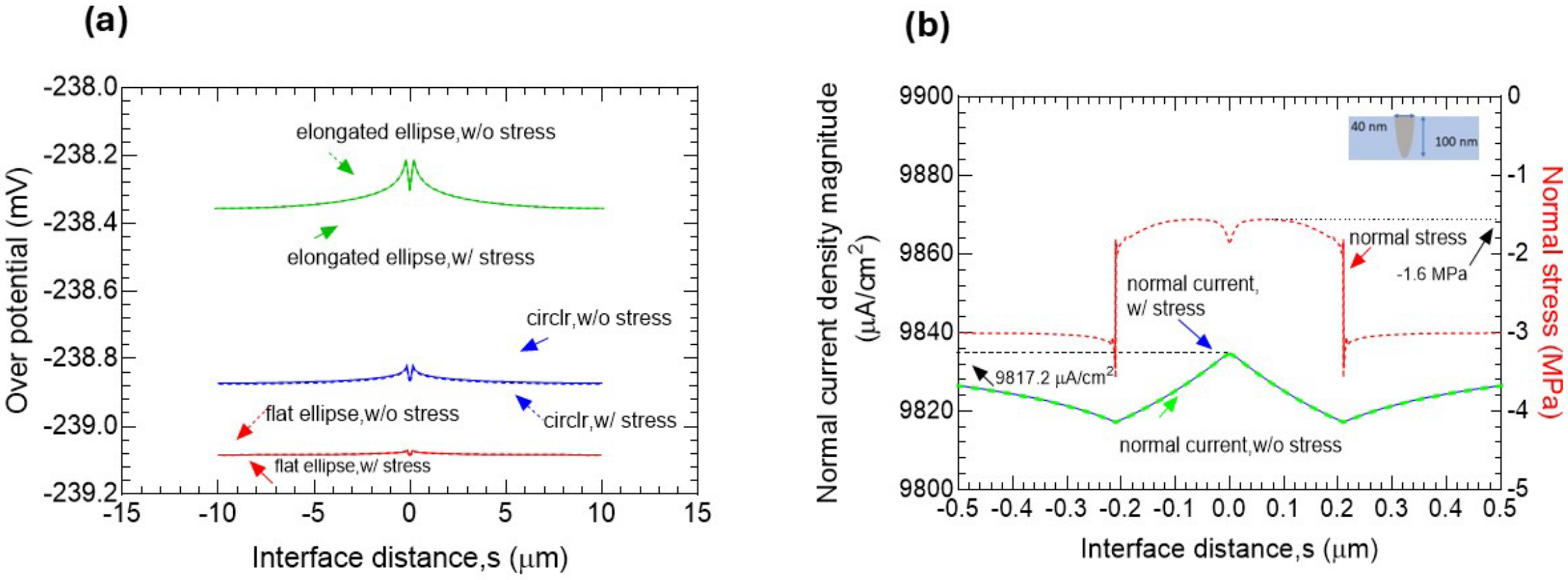

**Fig. S6 (a)** Interface over potential distribution along the roughness for different geometries for 100× the applied current density for plating condition and **(b)** magnitude of normal current and normal stress distribution along the interface for 100× the applied current density for plating condition.

## S6. Influence of Exchange Current Density Variations, Stripping

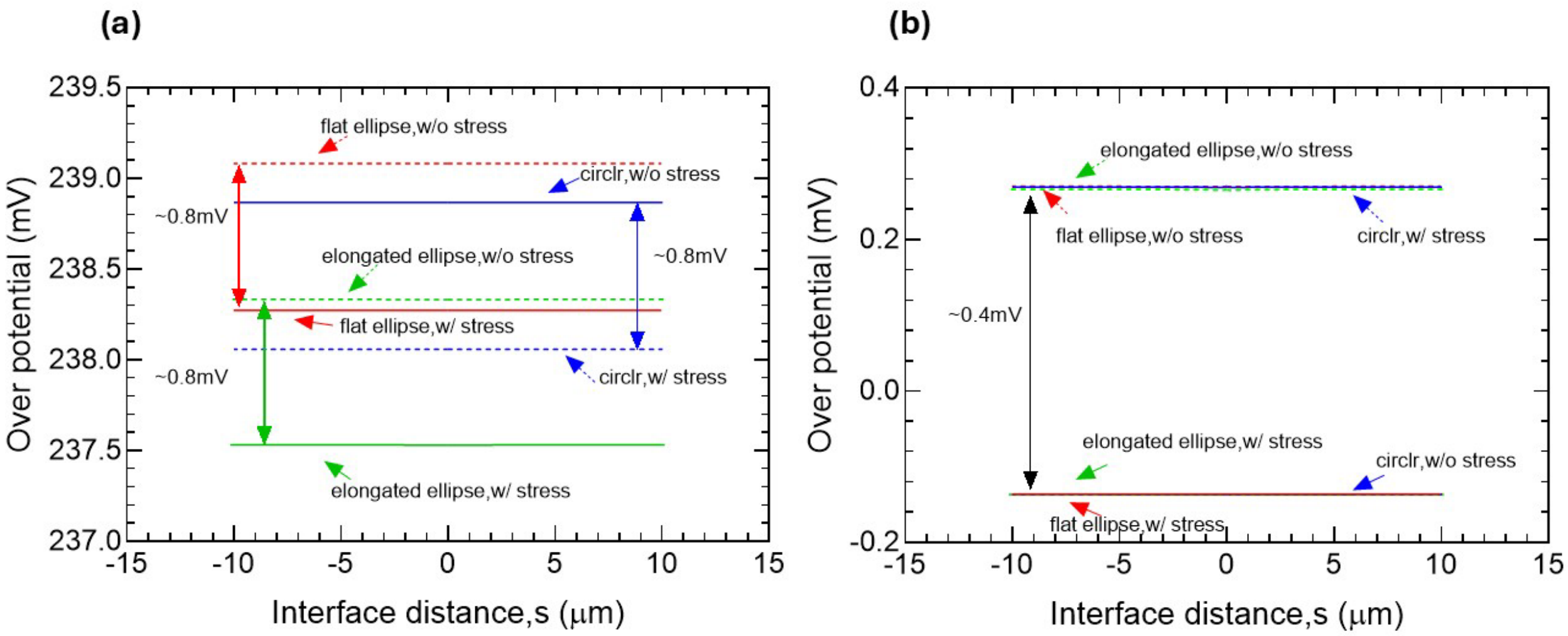

**Fig. S7** Interface overpotential distribution along the roughness for different geometries **(a)** 0.01× the experimental exchange current density; **(b)** 100× the experimental exchange current density for stripping condition.

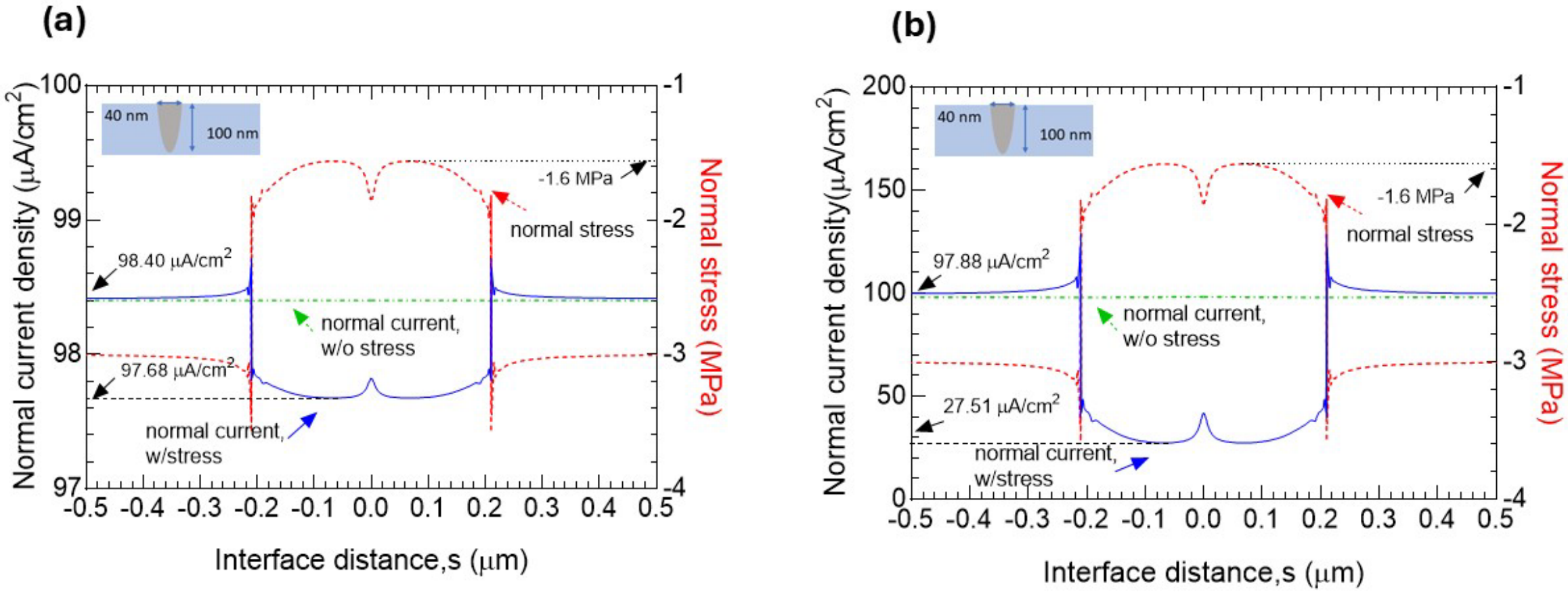

**Fig. S8** Magnitude of normal current and normal stress distribution along the interface **(a)** 0.01× the experimental exchange current density; **(b)** 100× the experimental exchange current density for stripping condition.

## S7. Influence of Applied Current Density Variations, Stripping

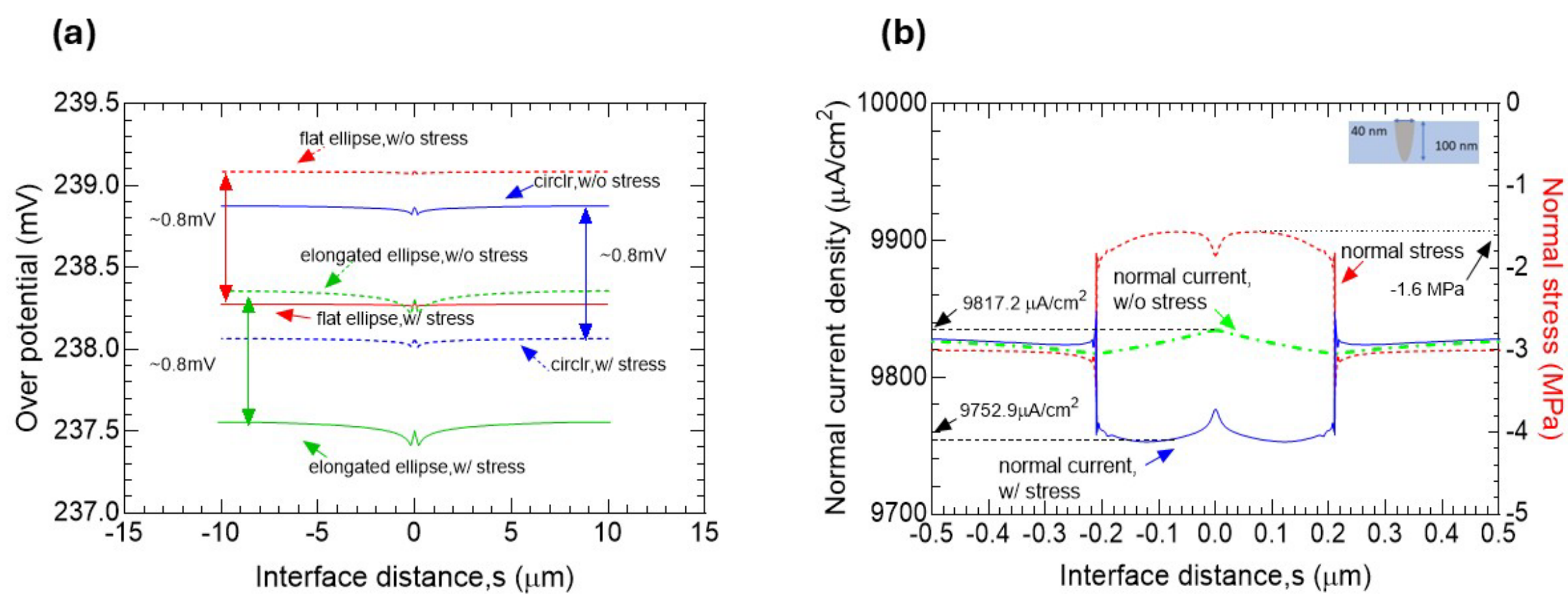

**Fig. S9 (a)** Interface over potential distribution along the roughness for different geometries for 100× the applied current density for stripping condition and **(b)** magnitude of normal current and normal stress distribution along the interface for 100× the applied current density for stripping condition.

## S8. Evaluation of Effective Ionic Conductivity

The purpose of this section is to analytically derive the effective ionic conductivity for the model incorporating a protective $Li_3N$ layer. The derivation is carried out for a smooth, flat Li/SE interface with a uniform protective layer, corresponding to the geometry shown in Fig. 2(d) in the absence of interfacial surface roughness. Since the anodic current is positive in the Butler-Volmer equation, $i_{BV}$ for plating is negative and the magnitude is identical to $i_{app}$ for a smooth flat surface. In this case, interfacial normal stress $(\sigma_n)$ is identical to the appled compressive stress.

Eqs (S1) and (S2) are obtained by solving Eqs. (1) and (3) subject to the boundary conditions specified in Eq. (4). These expressions provide an analytical basis for evaluating effective conductivity.

$$\phi_{SE} = \frac{i_{app}}{\sigma_{SE}^{+}}(L-y) + \frac{i_{app}}{\sigma_{layer}^{+}} t - i_{BV_layer} R_{\mathrm{int_layer}} - \phi_{eq,layer} \tag{S1}$$

$$\phi_{layer} = \frac{i_{app}}{\sigma_{layer}^{+}}(L+t-y) - i_{BV_layer} R_{\mathrm{int_layer}} - \phi_{eq,layer} \tag{S2}$$

Ionic conductivity plays a crucial role in charge-transfer kinetics in ASSBs. As shown in Table-I, the conductivity of $Li_3N$ layer is much less than the SE, and this could affect the ion transfer during cycling of the battery. To understand the relation of the ionic conductivity between the SE and the protective layer the following relation (Eq. S4) has been derived. Evaluating Eq. (S2) at $y$=0 and Eq. (S3) at $y=L+t$, we have

$$\phi_{layer}|_{y=L+t} - \phi_{SE}|_{y=0} = -i_{app}\left(\frac{\sigma_{layer}^{+} L + \sigma_{SE}^{+} t}{\sigma_{SE}^{+}\sigma_{layer}^{+}}\right) = -i_{app}\left(\frac{L+t}{\sigma_{eff}^{+}}\right) \tag{S3}$$

where $\sigma_{eff}^{+}$ is the effective ionic conductivity and expressed as

$$\frac{\sigma_{eff}^{+}}{\sigma_{SE}^{+}} = \frac{\dfrac{\sigma_{layer}^{+}}{\sigma_{SE}^{+}}}{\dfrac{t}{L+t}\left(1-\dfrac{\sigma_{layer}^{+}}{\sigma_{SE}^{+}}\right)+\dfrac{\sigma_{layer}^{+}}{\sigma_{SE}^{+}}} = \frac{\alpha}{x(1-\alpha)+\alpha} \tag{S4}$$

here $x = \frac{t}{L+t}$ and $\alpha = \frac{\sigma_{layer}^{+}}{\sigma_{SE}^{+}}$